\documentclass[10pt]{article}
\usepackage{graphicx}%
\usepackage{siunitx}
\usepackage{bm}
\usepackage[caption=false]{subfig}
\usepackage{mathtools}
\usepackage{multirow}
\usepackage{amssymb}
\usepackage{braket}
\usepackage{amsmath}
\usepackage{placeins}
\usepackage{cancel}
\usepackage{xcolor}
\usepackage{spreadtab}
\usepackage{dsfont}

\usepackage[numbers]{natbib}
\usepackage[margin=1in]{geometry}
\usepackage{authblk}

\newcommand\kf{K} 
\newcommand\twokf{Q} 

\title{\textbf{Importance of Non-Adiabatic Effects on Kohn Anomalies in 1D metals}}

\author[1]{Enrico Marazzi}
\author[2,3]{Samuel Poncé}
\author[1]{Jean-Christophe Charlier}
\author[2,3,4]{Gian-Marco Rignanese}

\affil[1]{Institute of Condensed Matter and Nanosciences, Université catholique de Louvain, Chemin des Étoiles 8, B-1348 Louvain-la-Neuve, Belgium.}
\affil[2]{European Theoretical Spectroscopy Facility, Institute of Condensed Matter and Nanosciences, Université catholique de Louvain, Chemin des Étoiles 8, B-1348 Louvain-la-Neuve, Belgium.}
\affil[3]{WEL Research Institute, Avenue Pasteur, 6, 1300 Wavre, Belgium.}	
\affil[4]{School of Materials Science and Engineering, Northwestern Polytechnical University, No. 127 Youyi West Road, Xi’an 710072 Shaanxi, P. R. China.}

\begin{document}

\maketitle

\begin{abstract}
Kohn anomalies are kinks or dips in phonon dispersions which are pronounced in low-dimensional materials.  
We investigate the effects of non-adiabatic phonon self-energy on Kohn anomalies in one-dimensional metals by developing a model that analyzes how the adiabatic phonon frequency, electron effective mass, and electron-phonon coupling strength influence phonon mode renormalization.
We introduce an electron-phonon coupling strength threshold for low-temperature system instability, providing experimentalists with a tool to predict them.
Finally, we validate the predictions of our model against first-principles calculations on a 4\,\r{A}-diameter carbon nanotube. 
\end{abstract}

Kohn anomalies (KAs) are softening of phonon modes in particular portions of the Brillouin zone (BZ)~\cite{kohn1959}. 
In Kohn's seminal work, the origin of this phenomenon was attributed to the screening of lattice vibrations by the excitation of electrons at the Fermi surface (FS). 
Soon after this theoretical prediction, KAs were observed by inelastic neutron scattering in lead~\cite{brockhouse1961}. 
Subsequently, they were seen in other metals~\cite{brockhouse1962,nakagawa1963,koenig1964}, superconductors~\cite{baron2004,aynajin2008}, and various other materials~\cite{powell1968,kulda2002}.
For a long time, the common belief was that electron and lattice dynamics were independent and that phonon instabilities were due solely to electronic effects~\cite{peierls1955}. 
However, this is only true for pure one-dimensional (1D) metals where the electronic response function diverges for the $\textbf{q}$ vectors connecting points in the FS, called perfect FS nesting. 
In quasi-1D, two-dimensional (2D), and three-dimensional (3D) materials, the electronic response function does not exhibit such divergence~\cite{zhu2015}.
For 2D and 3D systems, the situation is complicated by the fact that phonon softening occurs at $\textbf{q}$ vectors which are incommensurate with perfect FS nesting vectors~\cite{johannes2008}. 
It was therefore suggested that the origin of KAs lies in the electron-phonon coupling (EPC), since electronic and lattice instabilities always occur simultaneously. 
FS nesting contributes to creating instabilities, but is not the only driving force.

In this framework, recent computational efforts have focused on improving the accuracy of phonon dispersion in metals. 
One method incorporates anharmonic corrections to the phonon dispersion relations~\cite{tidholm2020}, while another accounts for non-adiabatic (NA) effects due to EPC~\cite{berges2023,caruso2017,girotto2023}, supporting the idea that EPC lies at the heart of these instabilities. 
However, the impact of NA effects on 1D systems remains unclear, which complicates the assessment of phonon mode softening.
Another open question is to identify the physical conditions that drive the instability.

In this work, we develop a 1D model to show that a minimum EPC strength is required to yield an imaginary phonon frequency.
This threshold depends only on the adiabatic phonon frequency $\omega_{\textbf{q}\nu}$ and the electron effective mass $m^*$, which decreases as $\omega_{\textbf{q}\nu}$ decreases and $m^*$ increases.
As such, the threshold provides a simple predictive tool for identifying low-temperature structural instabilities experimentally. 
Finally, we perform first-principles calculations on a 4~\r{A}-diameter carbon nanotube and find good agreement with our model.

In first-principles computations, phonon frequencies are usually calculated using density functional perturbation theory (DFPT)~\cite{gonze1997,gonze1997a} within the adiabatic approximation. 
They are obtained by diagonalizing the DFPT dynamical matrix $D^\text{DFPT}$, which is the sum of the bare dynamical matrix $D^\text{b}$ and the static-adiabatic phonon self-energy $\Pi^\text{A}$~\cite{berges2023}:
\begin{equation}
    D_{\textbf{q}\nu\mu}^\text{DFPT}(\omega=0,\sigma) = D_{\textbf{q}\nu\mu}^\text{b} + \Pi_{\textbf{q}\nu\mu}^\text{A}(\omega=0,\sigma),
    \label{eq:DFPT_DM}
\end{equation}
where $\textbf{q}$ is the phonon wavevector, $\nu$ and $\mu$ are phonon modes, $\omega$ is the vibrational dynamical variable and $\sigma$ is a large electronic temperature, which acts as a smearing parameter.
Going beyond DFPT to obtain the dynamical matrix $D$ requires to compute the non-adiabatic phonon self-energy $\Pi^\text{NA}$:
\begin{equation}
    D_{\textbf{q}\nu\mu}(\omega,T) = D_{\textbf{q}\nu\mu}^\text{b} + \Pi_{\textbf{q}\nu\mu}^\text{NA}(\omega,T),
    \label{eq:full_DM}
\end{equation}
where $T$ is the electronic temperature.
The bare dynamical matrix can be difficult to compute in a pseudopotential \emph{ab-initio} framework~\cite{berges2023} such that 
it is advantageous to compute $D$ from $D^\text{DFPT}$ by subtracting the static-adiabatic phonon self-energy and adding the dynamical non-adiabatic one: 
\begin{equation}
   \!\! D_{\textbf{q}\nu\mu}(\omega,T) = D_{\textbf{q}\nu\mu}^\text{DFPT}(0,\sigma) - \Pi_{\textbf{q}\nu\mu}^\text{A}(0,\sigma) \!+\! \Pi_{\textbf{q}\nu\mu}^\text{NA}(\omega,T).
    \label{eq:DM}
\end{equation}
The dynamical NA phonon self-energy can be computed with two screened vertices as~\cite{Calandra2010,giustino2017,berges2023,Marini2025,Stefanucci2025}:
\begin{equation}
    \Pi_{\textbf{q}\nu\mu}^{\text{NA}}(\omega,T) = \frac{2}{\hbar} \sum_{mn} \int \frac{d\textbf{k}}{\Omega^\text{BZ}} \\
    \times   \left[g^{mn\mu}_{\mathbf{k}\mathbf{q}}(0,\sigma) \right]^*
    \chi^{mn}_{\mathbf{k}\mathbf{q}}(\omega,T)g^{mn\nu}_{\mathbf{k}\mathbf{q}}(0,\sigma), 
    \label{eq:Pi_dynamic}
\end{equation}
where $g^{mn\nu}_{\mathbf{k}\mathbf{q}}(0,\sigma)$ are the screened EPC matrix elements,  $\Omega^{\text{BZ}}$ is the Brillouin-Zone volume, $m$ and $n$ are electron band indices, $\mathbf{k}$ is an electron wavevector, and 
$\chi^{mn}_{\mathbf{k}\mathbf{q}}(\omega,T)$ is bare electronic susceptibility given by:
\begin{equation}
\chi^{mn}_{\mathbf{k}\mathbf{q}}(\omega,T) = \bigg[\frac{f_{m\textbf{k}+\textbf{q}}(T)-f_{n\textbf{k}}(T)}{\varepsilon_{m\textbf{k}+\textbf{q}}-\varepsilon_{n\textbf{k}}-\hbar(\omega+i\eta)}\bigg],
\label{eq:chi}
\end{equation}
where $f$ is the Fermi-Dirac (FD) distribution function, $\varepsilon_{n\mathbf{k}}$ the DFT electron eigenvalues, and $\eta$ an infinitesimal quantity.

Here, we develop a 1D model with a parabolic electronic band $\varepsilon_k$ crossing the FS and a single-phonon mode $\nu$ aiming to study the behavior of the phonon self-energy due to EPC.
More details about this model can be found in Sect.~\ref{secSI:model} of the Supplementary Material~\cite{SI}. 
In this model, we assume a constant EPC $|g_{\twokf}|$ and a parabolic dispersion $\varepsilon_k\!=\! \hbar^2(k^2\!-\!\kf^2)/2m^*$.
Here, $\kf\!=\!k_\mathrm{F}$ is the Fermi wavevector that establishes the KA at $q\!=\!\twokf\!=\!2\kf$. 
In the model, we exclude the indices $n$, $m$, and $\nu$ when they are not pertinent, for the sake of clarity.
The assumption of a constant $|g_{\twokf}|$ is justified because the temperature dependence in Eq.~\eqref{eq:Pi_dynamic} arises only from the FD distribution in Eq.~\eqref{eq:chi}.
Therefore, this dependence has a contribution only from a small region around the Fermi level, where the EPC can be approximated as a constant.
Since this approximation worsens at higher temperature, we focus on the low-temperature regime.

Using first-order perturbation theory, the NA correction results in a renormalization of the phonon frequency which can be evaluated by retaining the diagonal elements of the self-energy $\Pi^{\mathrm{NA}}$ only:
\begin{equation}
    \Omega_{\twokf}^2(T) = 
    \omega_{\twokf}^2 \!+\! 
    \gamma_{\twokf}^2(\Omega_{\twokf}\!-\!i\gamma_{\twokf},T) \!+\! 
    2 \omega_{\twokf} 
    \times
    \!\Bigl[\mathrm{Re}\Pi^\text{NA}_{\twokf}(\Omega_{\twokf}\!-\!i\gamma_{\twokf},T) \!-\! \mathrm{Re}\Pi^\text{A}_{\twokf}(0,\sigma)\Bigr],
    \label{eq:O_renorm_tot}
\end{equation}
where $\Omega_{\twokf}$ is the renormalized phonon frequency, $\omega_{\twokf}$ the high-temperature adiabatic frequency that acts as the DFPT one and, $\gamma_{\twokf}$ the phonon linewidth due to the EPC that can be expressed as:
\begin{equation}
    \gamma_{\twokf}(\Omega_{\twokf}\!-\!i\gamma_{\twokf},T) = \\
    - \frac{\omega_{\twokf}}{\Omega_{\twokf}} \mathrm{Im}\Pi_{\twokf}^\textrm{NA}(\Omega_{\twokf}\!-\!i\gamma_{\twokf},T).
    \label{eq:gamma}
\end{equation}
Equations~\eqref{eq:O_renorm_tot} and \eqref{eq:gamma} are solved self-consistently 
and we choose an initial value $\gamma_{\twokf}$=1~meV.
Additional details are provided in Sect.~\ref{secSI:T0} of the Supplementary Material~\cite{SI}.

In Fig.~\ref{fig:Fig1}, we show the normalized solution to Eq.~\eqref{eq:O_renorm_tot}, defined as $\tilde{\Omega}_\twokf\equiv\frac{\Omega_\twokf}{\omega_\twokf}$, as a function of the normalized EPC matrix element:
\begin{equation}
    |\tilde{g}_\twokf| \equiv \frac{|g_\twokf|}{|g_\twokf|_0},
    \label{eq:ghat}
\end{equation}
where $|g_\twokf|_0$ is the value of $|g_\twokf|$ such that Eq.~\eqref{eq:O_renorm_tot} is verified for $\Omega_\twokf=0$.
The results presented in Fig.~\ref{fig:Fig1} are obtained at $T$=$0$~K for which the integral in Eq.~\eqref{eq:Pi_dynamic} can be computed analytically with a constant $|g_{\twokf}|$, as shown in Sect.~\ref{secSI:T0} of the Supplementary Material~\cite{SI}.
The smearing parameter is set to $\sigma=6000$~K, approximately corresponding to $40$~mRy, as this is a reasonable value used in DFT calculations for metals.
In Fig.~\ref{fig:Fig1}, different colors indicate different values of $\omega_\twokf$, ranging from 10 to 200 meV, while the shaded area in the inset represents variations of $m^*$ between 0.2 and 5.0. 
We limit the analysis to values of 0.2 $<m^{*}<$ 5.0 because perturbation theory is expected to fail when the bands become too flat or too dispersive. 
Additionally, we limit $\omega_\twokf$ to values smaller than 200 meV as this is an upper bound for phonon frequencies in materials.
The figure shows that the $\hat{\Omega}_\twokf$-$|\hat{g}_\twokf|$ relation closely follows a circular arc with a black dashed line.
In the inset, the deviation from the perfect circular shape is shown as a function of $|\hat{g}_Q|$ for different values of $\omega_\twokf$ and $m^*$. 
The shape of the $\hat{\Omega}_\twokf$-$|\hat{g}_\twokf|$ curve is affected by $\omega_\twokf$ and $m^*$ in two different ways: a larger $\omega_\twokf$ brings the curve closer to a circular arc, but amplifies the influence of the effective mass.
\begin{figure}
    \centering
    \includegraphics[width=0.55\linewidth]{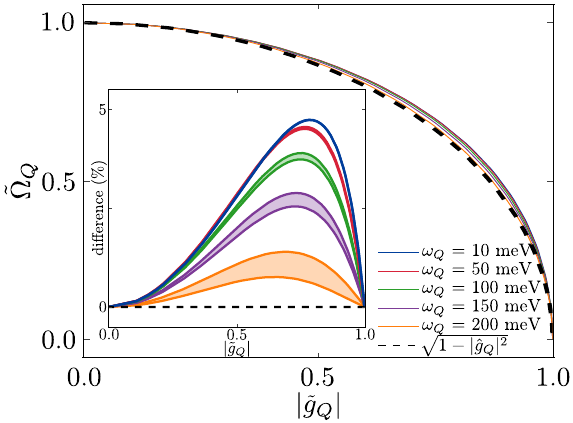}
    \caption{\label{fig:Fig1}Normalized phonon frequency $\Omega_\twokf$ as a function of the normalized EPC matrix element $|\hat{g}_\twokf|$, defined in Eq.~\eqref{eq:ghat} at 0~K. 
    Various values are reported for $\omega_\twokf$: 10, 50, 100, 150, and 200~meV in blue, red, green, purple, and orange respectively with $m^*=1$. 
    The black dashed line is the reference circular arc.
    The inset shows the deviation from this perfectly circular curve.
    The color code is the same and the shaded area account for different $m^*$ ranging between 0.2 and 5.0.
    }
\end{figure}

Since the value of $|g_\twokf|$ required to obtain $\Omega_\twokf\!=\!0$ remains finite even at $T=0$~K, we can consider $|g_\twokf|_0$ as a lower bound for the EPC vertices leading to system instability.
Therefore, being able to estimate the value of this threshold and compare it with the measured value of $|g_\twokf|$ is a tool to determine the stability of the system experimentally.
For this purpose, we approximate
\begin{equation}
    |g_\twokf|_0 \approx g_0 = \alpha + \beta \frac{\omega_\twokf^\delta}{(m^*)^\zeta},
    \label{eq:g0}
\end{equation}
with $\alpha$=7.278~meV, $\beta$=5.186~meV$^{1-\delta}$, $\delta$=0.763, and $\zeta$=0.518.
The parameters are computed by fitting Eq.~\eqref{eq:g0} on $|g_\twokf|_0$ for different values of $m^*$ and $\omega_\twokf$.
The approximation is validated in Sect.~\ref{secSI:g0} of the Supplementary Materials~\cite{SI}, where the percentage difference between $g_0$ and $|g_\twokf|_0$ is shown as a function of $\omega_\twokf$ and $m^*$.
Since $g_0$ depends only on $\omega_\twokf$ and $m^*$, it can be easily evaluated and compared with the experimental value of $|g_\twokf|$.

When $|g_{\twokf}|\!>\!g_0$, the renormalized frequency can become imaginary below a certain temperature.
In contrast, when $|g_{\twokf}|\!<\! g_0$, the frequency at $q=\twokf$ always remains positive even if a dip also appears in the renormalized bandstructure. 
This can be shown analytically at $T=\!$~0~K where the integral in Eq.~\eqref{eq:Pi_dynamic} yields~\cite{SI}:
\begin{equation}
    \int \mathrm{d}k \, \chi_{kq}(\Omega_{\twokf}\!- i\gamma_{\twokf},0)  = \frac{m^*}{2q}
    \times\!\log\!\!\left[
    \frac{
    \left(
    \Omega_{\twokf}^2 + \gamma_{\twokf}^2
    \right)^2
    }{
    \left(
    \!(\frac{q^2}{m^*}\!+\!\Omega_{\twokf})^2\!+\!\gamma_{\twokf}^2
    \right)\!\!
    \left(
    \!(\frac{q^2}{m^*}\!-\!\Omega_{\twokf})^2+\!\gamma_{\twokf}^2
    \right)
    } 
    \right]\!.
    \label{eq:chi0K}
\end{equation}
Its substitution into Eq.~\eqref{eq:O_renorm_tot} provides the value of $\Omega_{\twokf}(T\!=\!0)\!>\!0$.
For $T\!>\!$ 0~K, an analytical solution is no longer available for the integral of Eq.~\eqref{eq:Pi_dynamic} but it can be computed numerically.
In general, lower temperatures enhance the importance of the phonon frequency renormalization. 
However, as the temperature increases, $\Omega_\twokf$ first decreases, reaching a minimum before rising again to approach the adiabatic frequency at high temperature. 
Interestingly, when $|g_{\twokf}|\!\ll\!g_0$, $\Omega_{\twokf}$ reaches a minimum value at the temperature 
\begin{equation}
    T^{\rm min} = \frac{\omega_{\twokf}}{4.6k_B},
    \label{eq:Tmin}
\end{equation}
which does not depend on $m^*$ as shown in Sect.~\ref{secSI:tmin} of the Supplementary Material~\cite{SI}.
We also remark that when $|g_{\twokf}| \rightarrow g_0$, the renormalization effect is more pronounced.
Therefore, for any 1D system, $g_0$ constitutes a threshold for the EPC strength above which the system becomes unstable at low temperature. 
It depends solely on $\omega_\twokf$ and $m^*$ and Eq.~\eqref{eq:g0} suggests that the threshold increases with larger $\omega_\twokf$ and smaller $m^*$.
Therefore, we expect a system to be prone to instability if the adiabatic frequency is low and the effective mass large.

We evaluated the validity of our model by comparing it with first-principles calculations on the (3,3) carbon nanotube (CNT) and the boron and strained gold monoatomic chains.
We use \textsc{Quantum Espresso}~\cite{QE-2009,QE-2017,QE-2020} and \textsc{EPW}~\cite{Giustino2007,Ponce2016,Lee2023} to compute the DFT electron bandstructures, the DFPT phonon frequencies, the EPC matrix elements, and the phonon self-energies. 
The local density approximation (LDA) is used with a plane-wave basis set of 100~Ry and a norm-conserving pseudopotential from \textsc{Pseudo-Dojo}~\cite{vansetten2018}. 
The CNT is modeled in a hexagonal unit cell with the z-axis aligned with the tube axis, and a $13$~\r{A}-vacuum along the off-axis directions to prevent spurious interactions between periodic replicas.
Reciprocal space sampling is converged for each smearing temperature, with k-grids ranging from 1$\times$1$\times$10$^2$ to 1$\times$1$\times$10$^4$, and q-grids from 1$\times$1$\times$50 to 1$\times$1$\times$200. 
Electron bands and EPC matrix elements are interpolated using maximally localized Wannier functions~\cite{Marzari2012,w90} as implemented in \textsc{EPW}. 
Eq.~\eqref{eq:Pi_dynamic} is calculated using a broadening parameter $\eta\!=\!k_BT/100$ with uniform grids ranging from 1$\times$1$\times$10$^4$ for $T\!=\!1500$~K to 1$\!\times\!$1$\!\times\!$10$^6$ for $T\!=\!50$~K.

\begin{figure*}[ht]
    \centering
    \includegraphics[width=0.75\linewidth]{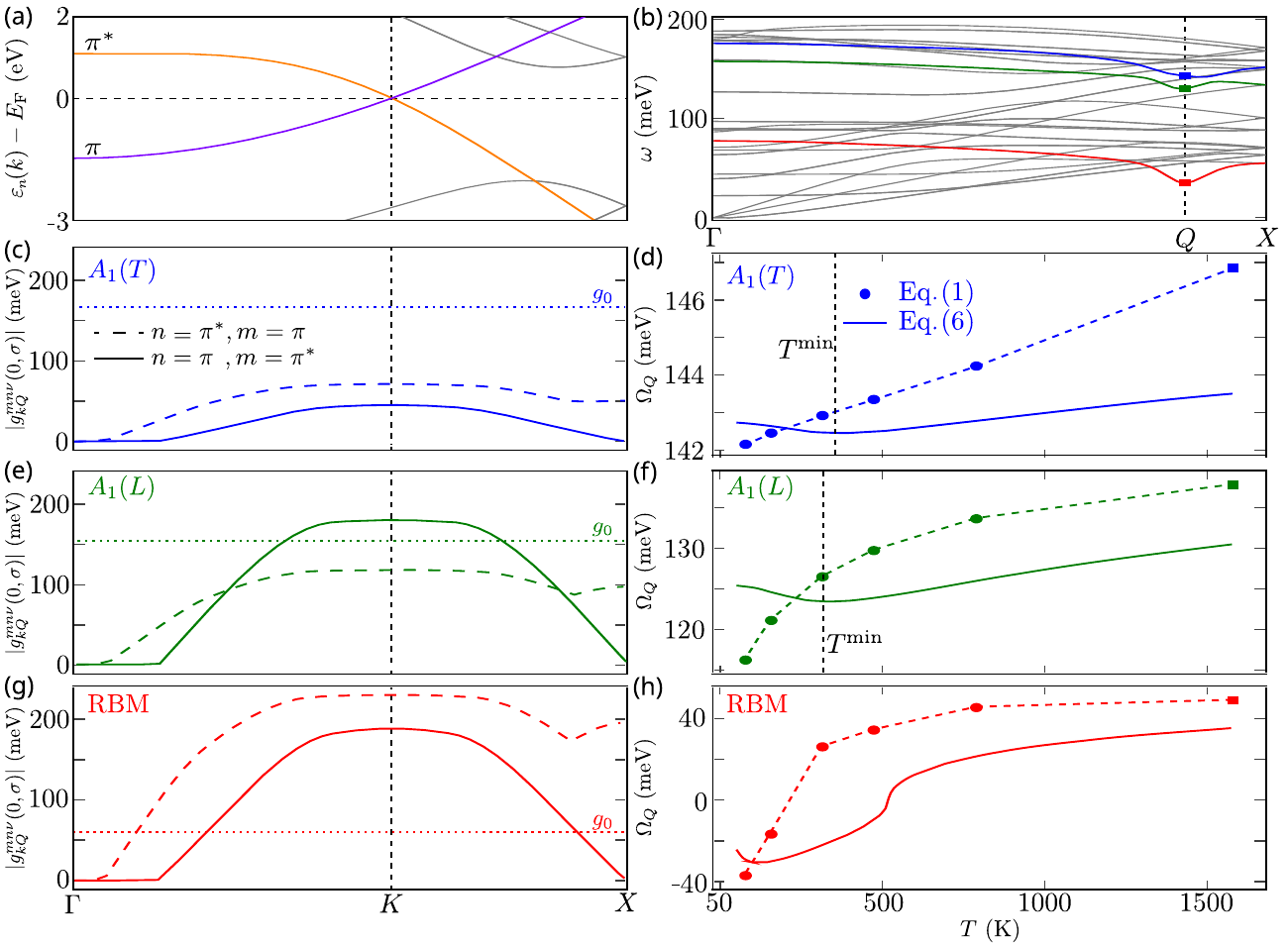}    \caption{\label{fig:cntb}The (a) electronic and (b) phonon bandstructures of the (3,3) carbon nanotube.
    The EPC matrix elements for $A_1(T)$ (c), $A_1(L)$ (e) and RBM (g).
    The dotted line is $g_0$ computed with Eq.~\eqref{eq:g0}.
    The renormalized phonon frequency for the three modes (d), (f) and (h).
    The solid line is computed with Eq.~\eqref{eq:O_renorm_tot} while the dashed line is the DFPT reference.
    }
\end{figure*}

The (3,3) CNT~\cite{connetable2005,barnett2005} is a one-dimensional metal with a Dirac cone at $\kf \simeq 0.575\Gamma$-X.
This band structure suggests a perfect FS nesting at $2(1-\kf) \simeq 0.85\Gamma$-X.
The electron and phonon bandstructures, computed with $\sigma=10$~meV and a cold smearing~\cite{marzari1999}, are shown in Fig.~\ref{fig:cntb}(a) and (b), respectively.
The phonon bandstructure shows a dip for three phonon modes at the expected position: the radial breathing mode (RBM) with an adiabatic $\omega_{\twokf} = 35.28$~meV, the in-plane longitudinal and transverse optical mode ($A_1(T)$ and $A_1(L)$) with $\omega_{\twokf} = 130.11$ and 143.42~meV, respectively.
Lowering the smearing parameter $\sigma$ as a proxy for the temperature in the DFPT calculation, we observe that for the RBM, the frequency becomes negative at $\sim$220~K.
For the $A_1(L)$ and $A_1(T)$ modes, the shape of the DFPT frequencies as a function of temperature suggests a transition to a negative frequency at low temperature but is not observed in the first-principles calculations.
This is probably not observed because the transition temperature obtained within this framework would be very low, making its computation out of reach. 
The DFPT frequencies, computed with a Fermi-Dirac smearing, are depicted in Fig.~\ref{fig:cntb}(d), (f), and (h), with a dotted line for the corresponding mode with $\sigma=T$.
The effective mass at the Fermi level is $m^*\approx2.22$ for the $\pi$ band and $m^*\approx1.82$ for the $\pi^*$ band.
Thus, we consider $m^*=2$ as the effective mass to compare with the model.
In Fig.~\ref{fig:cntb}(c), (e), and (g), the EPC matrix elements for the three modes involved in the renormalization are presented as a function of the wavevector $k$.
Let us first detail the predictions of our model based on the computed values of $g_0$ for each mode.
Using Eq.~\eqref{eq:g0}, 54.5~meV are found for the RBM, 129.0~meV for the $A_1(L)$ mode and 137.5~meV for the $A_1(T)$ mode, respectively.
Comparing these values with $|g^{nm\nu}_{\kf\twokf}|$, we conclude that: (i) the RBM is expected to soften as $|g^{ nm\rm RBM}_{\kf\twokf}|\!\gg\!g_0$, (ii) the $A_1(L)$ mode needs further investigation as $|g^{ nm\rm A_1(L)}_{\kf\twokf}|\!\sim\!g_0$ and (iii) the $A_1(T)$ mode should experience a negligible renormalization as $|g^{ nm\rm A_1(T)}_{\kf\twokf}|\!\ll\!g_0$.
Although $|g_{ k\twokf}^{ nm\nu}|$ assumes larger and smaller values than $g_0$, an appropriate comparison is to assess whether $|g_{ k\twokf}^{ nm\nu}|$ is larger or smaller than $g_0$ at the Fermi level, as detailed in Sect.~\ref{secSI:gkF2KF} of the Supplementary Material~\cite{SI}.
Furthermore, using Eq.~\eqref{eq:Tmin}, we predict that the renormalized frequency for the $A_1(T)$ mode should reach a minimum at 366~K.
In Figs.~\ref{fig:cntb}(d), (f) and (h), the renormalized phonon frequency at $q=\twokf$ are shown as a function of temperature with a solid line for the $A_1(T)$ mode, the $A_1(L)$ mode, and RBM, respectively.
Renormalized frequencies are computed using Eq.~\eqref{eq:O_renorm_tot} and $\omega_{\twokf}$ with a cold smearing~\cite{marzari1999}.
Firstly, as expected, the RBM softens, with a transition temperature of approximately 525~K, refining the estimates of Refs.~\cite{bohnen2004} and \cite{connetable2005}.
Next, we find that the $A_1(L)$ mode softens but remains positive throughout. 
The model of Eq.~\eqref{eq:Tmin} predicts that the phonon frequency should reach a minimum around 330~K and then rises to a finite value at low temperature.
Finally, as predicted, the $A_1(T)$ mode also softens but does not
become negative.
The temperature at which the renormalized frequency reaches a minimum is around 350~K, and the renormalization effect is even smaller.
These observations validate the predictions of our model.
Moreover, we can examine the effect of the adiabatic phonon frequency on the renormalization.
Since the effective mass at the Fermi level is constant in the system, a lower adiabatic frequency appears to promote instability.
Thus, while the RBM has a relatively low $g_0$ and does soften, the $A_1(L)$ and $A_1(T)$ modes have higher frequencies, leading to a larger $g_0$, which prevents softening.
Furthermore, since $|g^{ nm\rm RBM}_{\kf\twokf}| \gg g_0$, the transition occurs above room temperature, suggesting that the (3,3) CNT should experience a phase transition at room temperature in agreement with Ref.~\cite{connetable2005}.

To further validate our model, we computed the boron and strained gold chains, detailed in Sect.~\ref{secSI:DFT} of the Supplementary Materials~\cite{SI}, and further confirmed the results obtained from the theoretical model.
For the longitudinal mode of the boron chain, the DFPT phonon frequency is 138.15~meV, the electron effective mass 0.67 and $|g^{nm\rm LA}_{\kf\twokf}|\! = \!50$~meV.
Using the values of $\omega_{\twokf}$ and $m^*$ we obtain $g_0\!=\!240$~meV.
Consequently, we do not expect the renormalized phonon frequency to become negative.
Since $|g^{ nm\rm LA}_{\kf\twokf}|\!\ll\!g_0$, we also expect the renormalized frequency to show a minimum at a temperature $T^\mathrm{min}\!=\!350$~K.
For the longitudinal mode of the strained gold chain, the DFPT frequency is 5.19~meV.
The electron effective mass is 3, leading to $g_0\!=\!12.5$~meV.
Given that $|g^{ nm\rm LA}_{\kf\twokf}|\!\sim\!70$~meV, we expect the renormalized phonon frequency to become negative.
In both cases, Eq.~\eqref{eq:O_renorm_tot} confirms the predictions of the model.

In conclusion, we have developed a simple model to demonstrate how instabilities due to KA are promoted by a low $\omega_{\twokf}$ and a large $m^*$. 
Using these physical quantities, we identify a threshold for the EPC strength above which the system becomes unstable at sufficiently low temperature.
This threshold, $g_0$, can be compared with measured or computed EPC to easily estimate the stability of a system.
We validate the results of this model through first-principles calculations on the (3,3) CNT, which confirmed our theoretical predictions. 
This work emphasizes the dependence of Kohn anomalies on the intrinsic physical properties of one-dimensional systems and lays the groundwork for extending the model to two- and three-dimensional systems. 
Such extensions could offer insight into whether similar conditions govern KA behavior in higher-dimensional materials.

\subsection*{Acknowledgments}
E.M. acknowledges Victor Trinquet and Tom Van Waas for their assistance in solving the integrals in Section S3 of the Supplementary Material.
S.P. and G-M.R. acknowledge support from the Fonds de la Recherche Scientifique de Belgique (FRS-FNRS) and by the Walloon Region in the strategic axe FRFS-WEL-T.
J.-C.C. acknowledges financial support from the Fédération Wallonie-Bruxelles through the ARC project “DREAMS” (No. 21/26-116), from the EOS project “CONNECT” (No. 40007563), and from the Belgium FRS-FNRS through the research project (No. T.029.22F).
Computational resources have been provided by the supercomputing facilities of the Université catholique de Louvain (CISM/UCL) and the Consortium des Équipements de Calcul Intensif en Fédération Wallonie Bruxelles (CÉCI) funded by the Fond de la Recherche Scientifique de Belgique (F.R.S.-FNRS) under convention 2.5020.11 and by the Walloon Region.
The present research benefited from computational resources made available on Lucia, the Tier-1 supercomputer of the Walloon Region, infrastructure funded by the Walloon Region under the grant agreement n°1910247.

\clearpage

\section*{Supplementary Information}
\setcounter{section}{0}
\setcounter{figure}{0}
\setcounter{equation}{0}
\renewcommand{\thefigure}{S\arabic{figure}}
\renewcommand{\theequation}{S\arabic{equation}}
\renewcommand{\thesection}{S\arabic{section}}

\section{One-dimensional model}\label{secSI:model}

We develop a one-dimensional (1D) model to study the behavior of the phonon self-energy due to electron-phonon coupling (EPC).
We consider a one-band parabolic dispersion for the electron energies:
\begin{equation}
    \varepsilon_k = \frac{\hbar^2}{2m^*} \Bigl( k^2 - \kf^2\Bigr),
    \label{eq:ek}
\end{equation}
where k is a 1D wavevector, $m^*$ is the effective mass and $\kf = k_{\rm F}=\pi/2$ is the Fermi level position such that the perfect FS nesting vector is $q = \twokf = 2\kf = \pi$.
The equation for $\varepsilon_{k+q}$, where $q = \twokf$, imposing periodic boundary conditions in reciprocal space, is:
\begin{equation}
    \varepsilon_{k+\twokf} = 
    \begin{cases}
      \varepsilon^\mathrm{-}_{k+\twokf} =\frac{\hbar^2}{2m^*}\Bigl( (k+\twokf)^2 - \kf^2\Bigr) & \text{ k $\le$ 0}\\
      \varepsilon^\mathrm{+}_{k+\twokf} =\frac{\hbar^2}{2m^*}\Bigl( (k-\twokf)^2 - \kf^2\Bigr) & \text{ k $>$ 0}
    \end{cases}.
    \label{eq:ekq}
\end{equation}
For simplicity, we assume a single phonon mode and a constant EPC matrix element $|g_{\rm k\twokf}| = c$. 
In that case, it is possible to derive an analytic expression for the electronic susceptibility $\chi_{k\twokf}(\omega,T)$.
As reported in the main text, it can be expressed as:
\begin{equation}
    \chi_{k\twokf}(\omega,T) = \bigg[\frac{f_{k+\twokf}(T)-f_{k}(T)}{\varepsilon_{k+\twokf}-\varepsilon_{k}-\hbar(\omega_\twokf+i\eta)}\bigg],
    \label{eqSI:chi}
\end{equation}
where $\eta$ is an infinitesimal quantity and $f$ is the Fermi-Dirac distribution function:
\begin{equation}
    f_k(T)= \frac{1}{e^{\frac{\varepsilon_k}{k_BT}}+1},
    \label{eq:FD}
\end{equation}
where $\varepsilon_k$ is defined with the Fermi level at zero, $k_B$ is the Boltzmann constant, and $T$ as the temperature. 
The numerator of Eq.~\eqref{eqSI:chi} is:
\begin{equation}\label{eqSI:fd_diff1}
    f_{k+\twokf}(T)-f_k(T) = \frac{1}{e^{\frac{\varepsilon_{k+\twokf}}{k_BT}}+1} - \frac{1}{e^{\frac{\varepsilon_k}{k_BT}}+1}.
\end{equation}
Around the Fermi level ($k=\pm\kf$), the two dispersions can be written as a Taylor series with respect to $k$ as:
\begin{equation}
    \varepsilon_{k} = \frac{\hbar^2}{2m^*}\twokf(k-\kf) + \mathcal{O}(k^2)
\end{equation}
and
\begin{align}
    \varepsilon_{k+\twokf} =& 
    \begin{cases}
      \varepsilon^\mathrm{-}_{k+\twokf} = \ \ \frac{\hbar^2}{2m^*}\twokf(k+\kf) + \mathcal{O}(k^2) & \text{ k $\le$ 0}\\
      \varepsilon^\mathrm{+}_{k+\twokf} =-\frac{\hbar^2}{2m^*}\twokf(k-\kf) + \mathcal{O}(k^2) & \text{ k $>$ 0}
    \end{cases}.
    \label{eq:ekqt}
\end{align}

\begin{figure}
    \centering
    \includegraphics[width=0.5\textwidth]{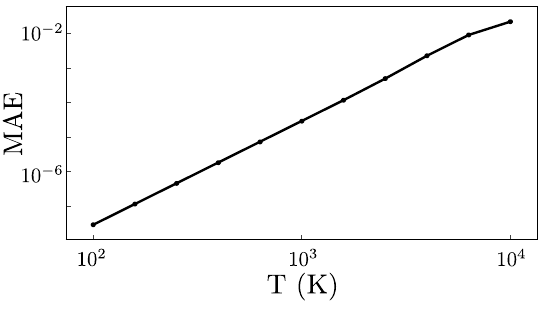}
    \caption{Mean absolute error of the difference between Eq.~\eqref{eqSI:fd_diff1} and Eq.~\eqref{eqsI:num-tanh}, across the  Brillouin Zone as a function of temperature for $m^*=1$.}
    \label{figSI:FigS1}
\end{figure}

We focus on the positive (+) sign and note that the (-) is symmetric, and we define:
\begin{equation}
\lambda \equiv -\frac{\hbar^2}{2m^*}\twokf(k-\kf),
\end{equation}
so that $\lambda \simeq \varepsilon^+_{k+\twokf} \simeq -\varepsilon_{k}$.
Knowing that $\frac{1}{e^{\theta}+1}-\frac{1}{e^{-\theta}+1} =  -\tanh(\frac{\theta}{2})$, Eq.~\eqref{eqSI:fd_diff1} thus becomes:
\begin{equation}
    f_{k+\twokf}^+(T)-f_k(T) = 
    \frac{1}{e^{\frac{\lambda}{k_{\rm B}T}+1}} - \frac{1}{e^{\frac{-\lambda}{k_{\rm B}T}+1}} =  -\tanh\frac{\lambda}{2k_{\rm B}T}.
    \label{eqsI:num-tanh}
\end{equation}
To assess the quality of the Taylor expansion, we show in Fig.~\ref{figSI:FigS1} the mean absolute error of the difference between Eq.~\eqref{eqSI:fd_diff1} and Eq.~\eqref{eqsI:num-tanh} across the entire Brillouin Zone (BZ) as a function of temperature. 
We find excellent agreement. 

The positive part of the denominator in Eq.~\eqref{eqSI:chi} is:
\begin{equation}
    \varepsilon^\mathrm{+}_{k+\twokf} - \varepsilon_k - (\omega_\twokf+i\eta)= 
    \frac{\hbar^2}{2m^*}\Bigl((k-\twokf)^2-\kf^2\Bigr) - \frac{\hbar^2}{2m^*}(k^2-\kf^2) - (\omega_\twokf+i\eta)= 2\lambda - (\omega_\twokf+i\eta).
\end{equation}
Combining the results, we obtain:
\begin{equation}
    \chi_{k\twokf}(\omega_\twokf,T) = - \frac{\tanh\frac{2\lambda}{4k_BT}}{2\lambda-\omega_\twokf-i\eta},
    \label{eqSI:chi_tanh}
\end{equation}
where
\begin{align} \label{eqSI:chi_tanh_real}
    {\mathrm {Re} }\chi_{k\twokf}(\omega_\twokf,T) =& -\frac{2\lambda-\omega_\twokf}{(2\lambda-\omega_\twokf)^2+\eta^2}\tanh\frac{2\lambda}{4k_BT} \\
       {\mathrm {Im} }\chi_{k\twokf}(\omega_\twokf,T) =& -\frac{\eta}{(2\lambda-\omega_\twokf)^2+\eta^2}\tanh\frac{2\lambda}{4k_BT}.
    \label{eqSI:chi_tanh_imag}
\end{align}  

These equations have been written for $k\!\ge\!0$, but since $\kf$ is at the middle of the irreducible BZ, $\chi_{k\twokf}$ at negative $k$ is symmetric and the whole integral is just the double of the integral of Eq.~\eqref{eqSI:chi_tanh}.

\section{Solution at T = 0~K}\label{secSI:T0}

We can solve exactly the integrals of Eq.~\eqref{eqSI:chi_tanh_real}
and \eqref{eqSI:chi_tanh_imag} when $T = 0$~K. 
At this temperature, the hyperbolic tangent is a Heaviside step function centered around $k=\kf$. 
As a result, the integral of Eq.~\eqref{eqSI:chi_tanh_real} becomes:
\begin{multline}
    \int_\mathrm{BZ} \mathrm{d}k\mathrm{Re
    }\chi_{k\twokf}(\omega_\twokf,0)  = 2 \Bigl(\int_0^\kf\frac{2\lambda-\omega_\twokf}{(2\lambda-\omega_\twokf)^2+\eta^2}dk -\int_\kf^\twokf \frac{2\lambda-\omega_\twokf}{(2\lambda-\omega_\twokf)^2+\eta^2}dk \Bigr) = \\
    = \frac{Q}{2E_Q} \log\frac{(\omega_\twokf^2+\eta^2)^2}{[(\omega_\twokf-E_Q)^2+\eta^2][(\omega_\twokf+E_Q)^2+\eta^2]},
    \label{eqSI:int_chi_real}
\end{multline}
where $E_Q = \frac{\hbar^2\twokf^2}{2m^*}$, while the integral of Eq.~\eqref{eqSI:chi_tanh_imag} becomes:
\begin{multline}
    \int_\mathrm{BZ} \mathrm{d}k\mathrm{Im}\chi_{k\twokf}(\omega_\twokf,0)  = 2
    \Bigl(\int_0^{\kf} \frac{\eta}{(2\lambda-\omega_\twokf)^2+\eta^2}dk -\int_\kf^\twokf \frac{\eta}{(2\lambda-\omega_\twokf)^2+\eta^2}dk \Bigr) = \\
    = \frac{Q}{E_Q}\Bigl(2\arctan\frac{\omega_\twokf}{\eta} -\arctan\frac{E_Q+\omega_\twokf}{\eta}-\arctan\frac{-E_Q+\omega_\twokf}{\eta}\Bigr).
    \label{eqSI:int_chi_imag}
\end{multline}
\begin{figure}
    \centering
    \includegraphics[width=0.5\linewidth]{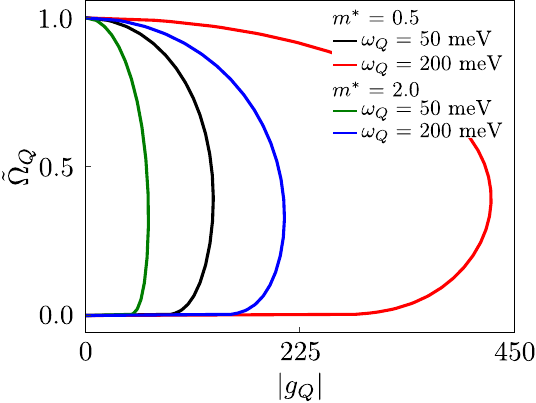}
    \caption{Solution of Eq.~\eqref{eq:O_renorm_tot} in the main manuscript as a function of $|g_{\rm \twokf}|$ considering the real part only at $T=0$~K. Therefore, here, $\gamma_{\rm \twokf}=0$.}
    \label{figSI:W-gREAL}
\end{figure}
With these results, we can solve Eq.~\eqref{eq:O_renorm_tot} of the main text, at first, by considering only the real part of the phonon frequency and of the non-adiabatic (NA) phonon self-energy.
For the electronic temperature, we consider $\sigma = 6000$~K, which is a typical value used in DFPT calculations.
The corresponding results are shown in Fig.~\ref{figSI:W-gREAL}.
We note that, as Eq.~\eqref{eqSI:int_chi_real} diverges at $\omega_\twokf=0$, all the metallic nanowires should be unstable.
However, we find two solutions for every value of $|g_{\twokf}|$, which does not seem physically possible.
To investigate this further, we consider the more general case where the solution of Eq.~\eqref{eq:O_renorm_tot} is the complex quantity $\Omega_{\rm \twokf}-i\gamma_{\rm \twokf}$. 
Here, the real part corresponds to the phonon frequency, while the imaginary part represents the phonon linewidth due to the EPC.
Using this formulation the integral in Eq.~\eqref{eqSI:int_chi_real} becomes:
\begin{multline}
    \int_\mathrm{BZ} \mathrm{d}k \mathrm{Re}\chi_{k\twokf}(\Omega_\twokf-i\gamma_\twokf,0) = \\
    2 \Bigl(\int_0^{\kf} \frac{2\lambda-\Omega_\twokf}{(2\lambda-\Omega_\twokf)^2+(\eta-\gamma_\twokf)^2}dk -\int_{\kf}^\twokf \frac{2\lambda-\Omega_\twokf}{(2\lambda-\Omega_\twokf)^2+(\eta-\gamma_\twokf)^2}dk \Bigr) = \\
    = \frac{\twokf}{2E_\twokf} \log\frac{(\Omega_\twokf^2+(\eta-\gamma_\twokf)^2)^2}{[(\Omega_\twokf-E_\twokf)^2+(\eta-\gamma_\twokf)^2][(\Omega_\twokf+E_\twokf)^2+(\eta-\gamma_\twokf)^2]}.
    \label{eqSI:int_chi_real_compfreq} 
\end{multline}
The phonon linewidth $\gamma_\twokf$ can be computed as:
\begin{multline}
    \gamma_\twokf(\Omega_\twokf-i\gamma_\twokf,0) = -\frac{\omega_\twokf}{\Omega_\twokf}\mathrm{Im}\Pi_{\rm \twokf}^\mathrm{NA}(\Omega_\twokf-i\gamma_\twokf,0) = 
    -\frac{\omega_\twokf}{\Omega_\twokf}|g_{\rm \twokf}|^2\int_\mathrm{BZ}\mathrm{d}k\mathrm{Im}\chi_{k\twokf}(\Omega_\twokf-i\gamma_\twokf,0) = \\
    -2\frac{\omega_\twokf}{\Omega_\twokf} |g_{\rm \twokf}|^2  \Bigl(\int_0^{\kf} \frac{\eta-\gamma_\twokf}{(2\lambda-\Omega_\twokf)^2+(\eta-\gamma_\twokf)^2}dk -\int_{\kf}^\twokf \frac{\eta-\gamma_\twokf}{(2\lambda-\Omega_\twokf)^2+(\eta-\gamma_\twokf)^2}dk \Bigr) = \\
    = \frac{\omega_\twokf}{\Omega_\twokf} |g_{\rm \twokf}|^2\frac{\twokf}{E_\twokf}\Bigl(2\arctan\frac{\Omega_\twokf}{\eta-\gamma_\twokf} -\arctan\frac{E_\twokf+\Omega_\twokf}{\eta-\gamma_\twokf}-\arctan\frac{-E_\twokf+\Omega_\twokf}{\eta-\gamma_\twokf}\Bigr),
    \label{eqSI:gamma_compfreq}
\end{multline}
where, $\omega_\twokf$ denotes the high-temperature frequency. 
Here, we notice that for $\Omega_\twokf \xrightarrow[]{}0$ Eq.~\eqref{eqSI:int_chi_real_compfreq} tends to:
\begin{equation}
    \int_\mathrm{BZ} \mathrm{d}k \mathrm{Re}\chi_{k\twokf}(0-i\gamma_\twokf,0)  \xrightarrow[]{} \frac{\twokf}{2E_\twokf}\log \Bigl(\frac{\gamma_\twokf^2}{E_\twokf^2+\gamma_\twokf^2}\Bigr)^2 = \frac{Q}{E_\twokf}\log \Bigl(\frac{\gamma_\twokf^2}{E_\twokf^2+\gamma_\twokf^2}\Bigr).
    \label{eqSI:Rechi0}
\end{equation}
Therefore, the integral does not diverge provided that $\gamma_\twokf$ is neither zero nor infinite. 
To verify this, we compute Eq.~\eqref{eqSI:gamma_compfreq} for $\eta\xrightarrow[]{}0$, $T\xrightarrow{}0$ and, $\Omega_\twokf\xrightarrow[]{}0$ and we obtain:
\begin{align}
    \gamma_\twokf(0+i\gamma_\twokf,0) =& \lim_{\Omega_\twokf \to 0} \frac{\omega_\twokf}{\Omega_\twokf}\frac{\twokf|g_{\rm \twokf}|^2}{E_\twokf}\Bigl(2 \arctan \frac{\Omega_\twokf}{\gamma_\twokf} - \arctan\frac{E_\twokf+\Omega_\twokf}{\gamma_\twokf} - \arctan\frac{-E_\twokf+\Omega_\twokf}{\gamma_\twokf}\Bigr)   \nonumber \\
    =& \lim_{\Omega_\twokf \to 0} \frac{\twokf\omega_\twokf|g_{\rm \twokf}|^2}{E_\twokf} \Bigl( 2\frac{\arctan(\Omega_\twokf/\gamma_\twokf)}{\Omega_\twokf} - \frac{\arctan(E_\twokf/\gamma_\twokf)}{\Omega_\twokf} + \frac{\arctan(E_\twokf/\gamma_\twokf)}{\Omega_\twokf} \Bigr) \nonumber\\
    =& \lim_{\Omega_\twokf \to 0} \frac{\twokf\omega_\twokf|g_{\rm \twokf}|^2}{E_\twokf} \Bigl( 2\frac{\arctan(\Omega_\twokf/\gamma_\twokf)}{\Omega_\twokf}\Bigr).
    \label{eq:lim1}
\end{align}
Knowing that $\lim_{x \to 0} \frac{\arctan(x)}{x} = 1$, from Eq.~\eqref{eq:lim1}, we can write
\begin{equation}
\lim_{\Omega_\twokf \to 0} \frac{\arctan(\Omega_\twokf/\gamma_\twokf)}{\Omega_\twokf} = \frac{1}{\gamma_\twokf},
\end{equation}
and so Eq.~\eqref{eq:lim1} becomes:
\begin{equation}
    \gamma_\twokf(0+i\gamma_\twokf,0) = \frac{2\twokf\omega_\twokf|g_{\rm \twokf}|^2}{E_\twokf} \frac{1}{\gamma_\twokf}.
\end{equation}
Therefore:
\begin{equation}
    \gamma_\twokf(0+i\gamma_\twokf,0) = \sqrt{\frac{2\twokf\omega_\twokf}{E_\twokf}}|g_{\rm \twokf}|.
    \label{eqSI:gammana}
\end{equation}
As a consequence, Eq.~\eqref{eqSI:Rechi0} becomes:
\begin{equation}
    \int_\mathrm{BZ}dk \mathrm{Re}\chi_{k\twokf}(0-i\gamma_\twokf,0) = \frac{\twokf}{E_\twokf}\log\frac{2\omega_\twokf|g_{\rm \twokf}|^2}{\frac{E_\twokf^3}{\twokf}+2\omega_\twokf|g_{\rm \twokf}|^2}= -\frac{\twokf}{E_\twokf} \log\Bigl(1+\frac{E_\twokf^3}{2\twokf\omega_\twokf|g_{\rm \twokf}|^2}\Bigl),
    \label{eqSI:rechiana}
\end{equation}
which depends only on the parameter of the model and has a finite value.
To validate these analytical expressions, we compare them with numerical calculations.
In Fig.~\ref{figSI:ana}, the integral of the real part of the polarizability (panel (a)), and $\gamma_\twokf$ (panel (b)), are shown as a function of $\Omega_\twokf$ at $T$ = 0~K.
The numerical solutions (solid lines) closely match the analytical solutions (dots) of Eqs.~\eqref{eqSI:rechiana} and \eqref{eqSI:gammana}, demonstrating the accuracy of the approach.

As a consequence, Eq.~\eqref{eq:O_renorm_tot} of the main article becomes:
\begin{equation}
    \Sigma =\omega_{\rm \twokf}^2 + \gamma_\twokf^2-\gamma_\twokf^2\log\Bigl(1+\frac{E_\twokf^2}{\gamma_\twokf^2}\Bigr)-\frac{E_\twokf}{\twokf}\gamma_\twokf^2\int_\mathrm{BZ}dk\chi_{k\twokf}(0,\sigma) = 0
    \label{eqSI:O_renorm_tot_ana}
\end{equation}
To validate this equation and the results presented in the main manuscript, we generate a set of 5000 parameter couples ($m^*$, $\omega_{\rm \twokf}$). 
For each couple, the corresponding EPC matrix elements is computed as follows:
\begin{equation}
    |g_\twokf| = g_0 ( 1 \pm \epsilon),
\end{equation}
where $g_0$ is the threshold value obtained using Eq.~\eqref{eq:g0} in the main manuscript, $\epsilon$ is a random value between 0 and 0.1.
We then substitute these values into Eq.~\eqref{eqSI:O_renorm_tot_ana}. 
If $\Sigma$ is zero or negative, the value of $|g_{\rm \twokf}|$ is large enough to cause instability.
This implies that Eq.~\eqref{eq:O_renorm_tot}, with $\Omega_{\rm \twokf}=0$, can be solved at a temperature $T\ge0$~K.
In contrast, if $\Sigma$ is positive, the equation has no solution, regardless of the temperature, meaning that the instability does not occur. 
Thus, when $|g_{\rm \twokf}|>g_0$ and $\Sigma <0 $ or $|g_{\rm \twokf}|<g_0$ and $\Sigma > 0$, the two approaches agree.
In Fig.~\ref{figSI:compg0-ana}, the accuracy of the model, defined as the percentage of instances in which the two approaches agree, is shown. 
The yellow regions correspond to conditions for which the model predicts the correct behavior with 100\% accuracy, while darker shades indicate reduced agreement. 
The accuracy decreases with lower adiabatic frequency or higher effective mass, reflecting the reduced reliability near the boundaries of the parameter ranges. 
In addition, the model becomes less reliable when $\epsilon$ is below 0.01.

In conclusion, an accurate description of the phonon frequency renormalization in metallic nanowires requires the imaginary part of the phonon frequency.
The broadening due to the EPC prevents the divergence in the electronic polarizability that would otherwise arise when considering only electronic effects.
This refinement leads to a more realistic and physically consistent picture, supporting the conclusions presented in the main text.
\begin{figure}
    \centering
    \includegraphics[width=0.99\linewidth]{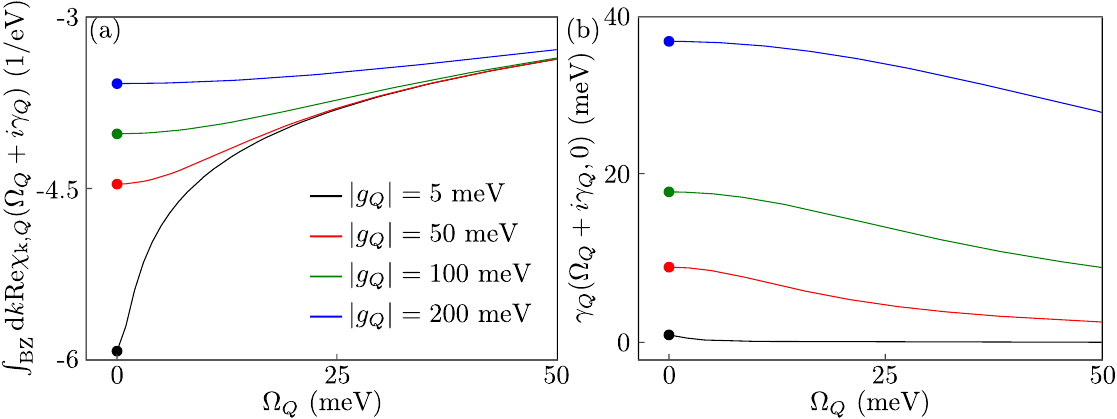}
    \caption{(a) Integral of the real part of the susceptibility at $T$ = 0~K as a function of the renormalized frequency $\Omega_\twokf$.
    (b) $\gamma_{\rm \twokf}$ at $T$ = 0~K as a function of the renormalized frequency $\Omega_\twokf$.
    In both panels, the lines are computed numerically, while the dots are the analytical results for $\Omega_\twokf = 0$ through Eqs.~\eqref{eqSI:rechiana} and \eqref{eqSI:gammana}, respectively.}
    \label{figSI:ana}
\end{figure}
\begin{figure}
    \centering
    \includegraphics[width=0.99\linewidth]{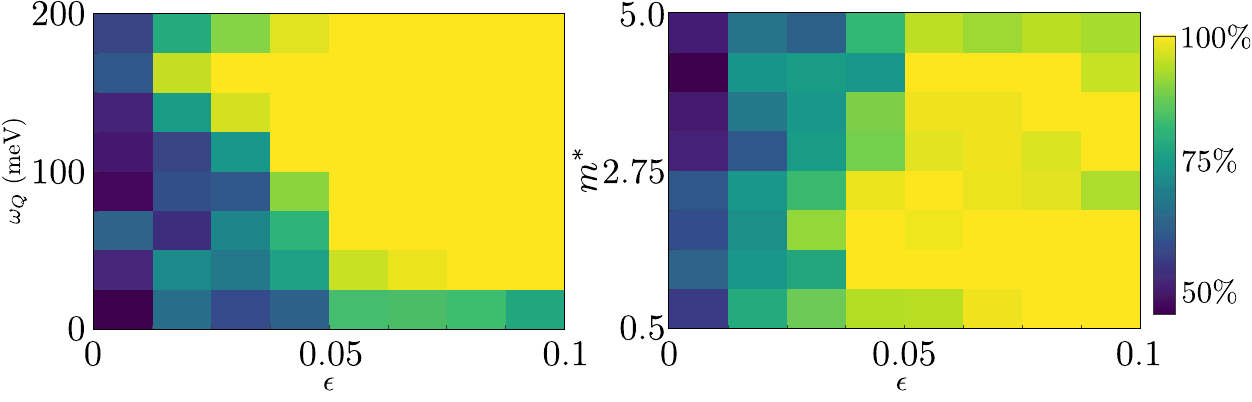}
    \caption{Comparison between the solution of Eq.~\eqref{eqSI:O_renorm_tot_ana} and the value of $g_0$ defined in Eq.~\eqref{eq:g0} of the main article for different sets of ($m^*$, $\omega_{\rm \twokf}$ and $|g_{\rm \twokf}|$). 
    The accuracy of the model is shown as a function $\epsilon$, and (a) frequency,  and (b) effective mass.
    }
    \label{figSI:compg0-ana}
\end{figure}
\newpage

\section{Validation of the $g_0$ model}\label{secSI:g0}
\begin{figure}
    \centering
    \includegraphics[width=0.5\linewidth]{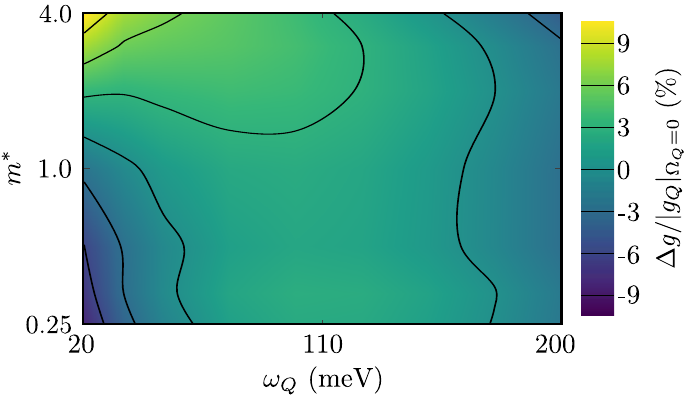}
    \caption{Percentage difference between $|g_\twokf|$ such that $\Omega_\twokf$ = 0 and the prediction of the model, $g_0$, as a function of $\omega_\twokf$ and $m^*$.}
    \label{figSI:map-g0-gQ}
\end{figure}
The approximation of $|g_\twokf|_0$ described in Eq.~\eqref{eq:g0} is validated in Fig.~\ref{figSI:map-g0-gQ}, where the percentage difference between $g_0$ and $|g_\twokf|_0$ is shown as a function of $m^*$ and $\omega_\twokf$. 
The largest discrepancies (up to $\pm10\%$) occur at the boundary of the $m^*$ and $\omega_\twokf$ parameter ranges, where the model is expected to be less accurate. 
These regions also correspond to those in which the model is less accurate; see Fig.~\ref{figSI:compg0-ana}.
Therefore, special care must be taken when $m^*$ is large, $\omega_\twokf$ is small, or $|g_\twokf|\approx g_0$, as the model becomes less reliable under these conditions. 

\begin{figure}
    \centering
    \includegraphics[width=0.99\textwidth]{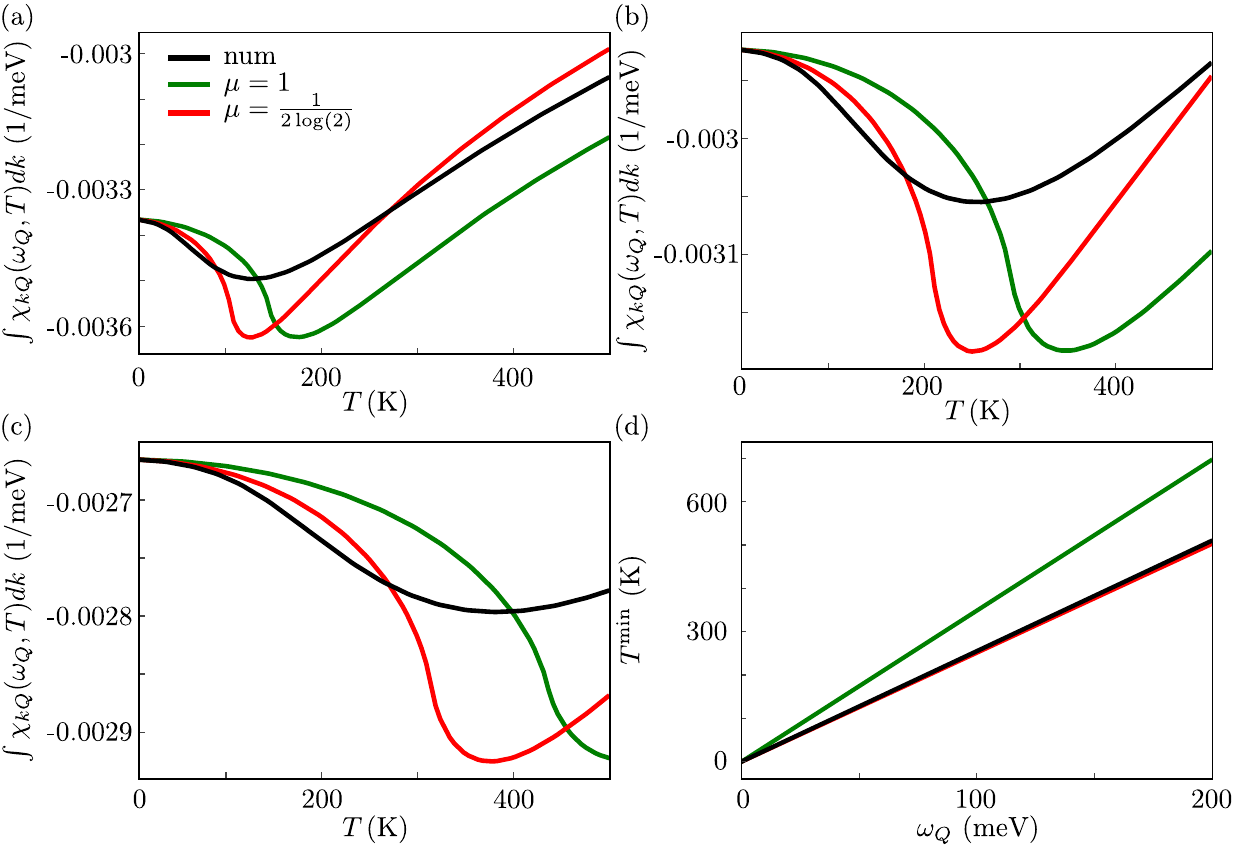}
    \caption{Evolution of the integral $\int_\mathrm{BZ}\mathrm{d}k\chi_\mathrm{k\twokf}$ as a function of temperature computed at $\omega_\twokf$ = (a) 50, (b) 100, and (c) 150 meV. 
    In panel (d), the temperature of the minimum is plotted as a function of $\omega_\twokf$. 
    In all the panels, the black line corresponds to the numerical solution, the green line to the linear approximation with $\mu=1$ and the red line with $\mu = \frac{1}{2\log(2)}$.
    } 
    \label{figSI:FigS3}
\end{figure}

\section{Temperature of the minimum}\label{secSI:tmin}

The function in Eq.~\eqref{eqSI:chi_tanh_real} is integrated and its minimum with respect to the temperature is found by solving for the points where its derivative equals zero.
Since the integral cannot be solved analytically, we adopt two approaches. 
First, we evaluate the integral numerically for the different parameter values.
Then, we use these numerical results to validate the analytical approximation, which is defined as follows:
\begin{equation}
    \tanh^{(1)}(x) \approx 
    \begin{cases}
       \text{sign}(x) & \text{  $|x|$ $>$ $\frac{1}{\mu}$}\\
       \mu x & \text{ $|x|$ $\le$ $\frac{1}{\mu}$}
    \end{cases}.  
    \label{eq:tanh1}                                                                                                                                                                                                                                                                                                                                                                                                                                                                    
\end{equation}
The integral becomes:
\begin{align}
    \int dk \chi_{k\twokf}^{(1)}(\omega_\twokf,T) = &
    2 \Bigl[ - \int_0^{\kf-\frac{2\twokf k_BT}{\mu E_\twokf}} \frac{2\lambda-\omega_\twokf}{(2\lambda-\omega_\twokf)^2+\eta^2} dk + \int_{\kf+\frac{2\twokf k_BT}{\mu E_\twokf}}^{\twokf} \frac{2\lambda-\omega_\twokf}{(2\lambda-\omega_\twokf)^2+\eta^2} dk \nonumber \\
    -& \int_{\kf-\frac{2\twokf k_BT}{\mu E_\twokf}}^{\kf+\frac{2\twokf k_BT}{\mu E_\twokf}} \frac{2\lambda-\omega_\twokf}{(2\lambda-\omega_\twokf)^2+\eta^2}\frac{2\mu\lambda}{4k_BT} dk\Bigr] \nonumber \\
    =&\frac{2\twokf}{E_\twokf} \Bigr[\log \frac{((\omega_\twokf+\frac{4k_BT}{\mu})^2+\eta^2)((\omega_\twokf-\frac{4k_BT}{\mu})^2+\eta^2)}{((\omega_\twokf+E_\twokf)^2+\eta^2)((\omega_\twokf-E_\twokf)^2+\eta^2)} \nonumber \\
    +&\frac{\mu \omega_\twokf}{4k_BT}\log\frac{(\omega_\twokf+\frac{4k_BT}{\mu})^2+\eta^2}{(\omega_\twokf-\frac{4k_BT}{\mu})^2+\eta^2}  \nonumber \\
    +&\frac{2\eta\mu}{4k_BT} \Bigl(\arctan\Bigl(\frac{\omega_\twokf+\frac{4k_BT}{\mu}}{\eta}\Bigr) - \arctan\Bigl(\frac{\omega_\twokf-\frac{4k_BT}{\mu}}{\eta}\Bigr)\Bigr) - 4\Bigr]
\end{align}
We can therefore compute analytically the position of the minimum as:
\begin{align}
    \frac{\partial \int\chi^{(1)}(\omega_\twokf,T)dk}{\partial T} =&  
    \frac{\twokf}{2E_\twokf} \Bigl[\frac{-2\frac{4k_B}{\mu}(\frac{4k_B}{\mu}T-\omega_\twokf)}{(\frac{4k_B}{\mu}T-\omega_\twokf)^2+\eta^2} + \frac{2\frac{4k_B}{\mu}(\frac{4k_B}{\mu}T+\omega_\twokf)}{(\omega_\twokf+\frac{4k_B}{\mu}T)^2+\eta^2}\nonumber \\
    +& \frac{\omega_\twokf\mu}{4k_BT}\frac{2\frac{4k_B}{\mu}(\omega_\twokf+\frac{4k_B}{\mu}T)((\omega_\twokf-\frac{4k_B}{\mu}T)^2+\eta^2)+2\frac{4k_B}{\mu}(\omega_\twokf-\frac{4k_B}{\mu}T)((\omega_\twokf+\frac{4k_B}{\mu}T)^2+\eta^2)}{((\omega_\twokf+\frac{4k_BT}{\mu})^2+\eta^2)((\omega_\twokf-\frac{4k_BT}{\mu})^2+\eta^2)} \nonumber \\
    -&\frac{\mu\omega_\twokf}{4k_BT^2} \log\frac{(\omega_\twokf+\frac{4k_B}{\mu}T)^2+\eta^2}{(\omega_\twokf-\frac{4k_B}{\mu}T)^2+\eta^2} + \frac{2\eta^2}{T} \Bigl(\frac{1}{(\omega_\twokf-\frac{4k_B}{\mu}T)^2+\eta} + \frac{1}{(\omega_\twokf+\frac{4k_B}{\mu}T)^2+\eta^2} \Bigr)\nonumber \\
    -& \frac{2\mu\eta}{4k_BT^2} \Bigl(\arctan\Bigl(\frac{\omega_\twokf+\frac{4k_B}{\mu}T}{\eta}\Bigr) - \arctan\Bigl(\frac{\omega_\twokf-\frac{4k_B}{\mu}T}{\eta}\Bigr)\Bigr]=0.
    \label{eqSI:dcdt_1st}
\end{align}
As $\eta$ is an infinitesimal quantity, as an approximation, we can consider it to be zero:
\begin{align}
    2&\Bigl(\frac{4k_B}{\mu}\Bigr)^2\Bigl(\frac{4k_B}{\mu}T-\omega_\twokf\Bigr)\Bigl(\omega_\twokf+\frac{4k_B}{\mu}T\Bigr)^2+2(\frac{4k_B}{\mu}\Bigr)^2\Bigl(\frac{4k_B}{\mu}T+\omega_\twokf\Bigr)\Bigl(\frac{4k_B}{\mu}T-\omega_\twokf\Bigr)^2\nonumber \\ 
    &+ 4 \omega_\twokf^2\Bigl(\frac{4k_B}{\mu}\Bigr)\Bigl(\omega_\twokf^2-\Bigl(\frac{4k_B}{\mu}\Bigr)^2\Bigr) -\omega_\twokf \Bigl(\omega_\twokf-\frac{4k_B}{\mu}T\Bigr)^2\Bigl(\omega_\twokf+\frac{4k_B}{\mu}T\Bigr)^2\log\frac{(\frac{4k_B}{\mu}T+\omega_\twokf)^2}{(\frac{4k_B}{\mu}T-\omega_\twokf)^2} \nonumber \\
    &=4\Bigl(\frac{4k_B}{\mu}T\Bigr)\Bigl(\omega_\twokf+\frac{4k_B}{\mu}T\Bigr)^2\Bigl(\frac{4k_B}{\mu}T-\omega_\twokf\Bigr)^2\Bigl[4\Bigl(\frac{4k_B}{\mu}T\Bigr)-\omega_\twokf\log\frac{(\frac{4k_B}{\mu}T+\omega_\twokf)^2}{(\frac{4k_B}{\mu}T-\omega_\twokf)^2} \Bigr] = 0.
\end{align}
We have a solution for $T=0$ K and another one that we can compute as:
\begin{equation}
    4\frac{4k_BT}{\mu\omega_\twokf}-\log\frac{(\frac{4k_BT}{\mu\omega_\twokf}+1)^2}{(\frac{4k_BT}{\mu\omega_\twokf}-1)^2}  = 0.
\end{equation}
Defining $x = \frac{4k_BT}{\mu\omega_\twokf}$, we get:
\begin{equation}
    4x-\log\frac{(x+1)^2}{(x-1)^2} = 0.
\end{equation}
The solution for this equation is $x \sim 1.199678 = \alpha$, and therefore, we obtain a linear relation of the form 
\begin{equation}
    T \sim \frac{\alpha\mu\omega_\twokf}{4k_B} \sim \frac{\mu\omega_\twokf}{3.33k_B}.
    \label{eqSI:Tmin_mu}
\end{equation}
This other solution is linearly dependent on $\omega_\twokf$ and $\mu$, but, interestingly does not depend on $m^*$. 
To estimate the value of $\mu$ that better approximate the hyperbolic tangent, we adopt two approaches: (i) $\mu=1$ as $\tanh(x)~=~x~+~\mathcal{O}(x^2)$ is the linear term of the Taylor expansion close to zero; (ii) we estimate the slope that minimize the integral of the difference between $\tanh^{(1)}(x)$ and $\tanh(x)$:
\begin{align}
    \Sigma =& \int_0^{\kappa} \tanh^{(1)}(x) - \tanh(x)dx = \int_0^{1/\mu} \mu x+\int_{1/\mu}^\kappa dx - \log(\cosh(\kappa)) \nonumber \\
    =& \frac{1}{2\mu} + \kappa - \frac{1}{\mu} -\log(\cosh(\kappa)) = - \frac{1}{2\mu} + \kappa - \log(\cosh(\kappa)).
\end{align}
Therefore, we can look for the value of $\mu$ that makes $\Sigma=0$ as $\kappa\xrightarrow{}\infty$.
\begin{equation}
    \frac{1}{\mu} = 2 \lim_{\kappa\xrightarrow[]{}0} \kappa - \log(\cosh(\kappa)) = 2 \log(2),
\end{equation}
where $\log(2)$ is the natural logarithm of 2.
Thus, we obtain $\mu = \frac{1}{2\log(2)}$.

In Fig. \ref{figSI:FigS3}, the integral is solved numerically and using the linear approximation with $\mu=1$ and $\frac{1}{2\log(2)}$  as a function of the temperature for three different frequencies $\omega_\twokf=$ 50, 100, and 150~meV in panels (a), (b) and (c), respectively. 
We notice that all the curves tend to the same value at zero temperature as both $\tanh$ and $\tanh^{(1)}$ tend to a Heaviside step function at $T=0$~K.
This is the local maximum that corresponds to the solution at $T=0$~K.
The other solution is the position of the minimum, which linearly depends on $\omega_\twokf$ and $\mu$.
In Fig.~\ref{figSI:FigS3}(d), the position of the minimum as a function of temperature is shown.
We notice that the linear approximation with $\mu=\frac{1}{2\log(2)}$ better reproduces the position of the minimum.
If we substitute the values of $\mu$ in Eq.~\eqref{eqSI:Tmin_mu}, we obtain a slope $\approx1/3.33k_B$ for $\mu=1$ and $\approx1/4.6k_B$ for $\mu=\frac{1}{2\log(2)}$.

This becomes relevant when the renormalization is small. 
If $\Omega_\twokf\approx\omega_\twokf$, Eq.~\eqref{eq:O_renorm_tot} does not need to be solved self-consistently, so the renormalization is larger at the position of the minimum.
This position can be determined by knowing the adiabatic phonon frequency through Eq.~\eqref{eqSI:Tmin_mu}.

\section{Predominance of the value of the EPC matrix element close to the Fermi level}\label{secSI:gkF2KF}

Throughout this work, we assume that $|g_{\rm \twokf}(0,\sigma)|$ is $k$-independent. 
However, to establish a connection between the model and realistic systems, where the EPC matrix element depends on $k$, we must extract an effective value from the computed one.
To achieve this, we model $|g_{\rm k\twokf}(0,\sigma)|$ in two different ways and show that $|g_{\rm k=\kf \twokf}(0,\sigma)|$ is the best value for comparison with the model to infer stability.

\begin{figure}[ht]
    \centering
    \includegraphics[width=0.99\linewidth]{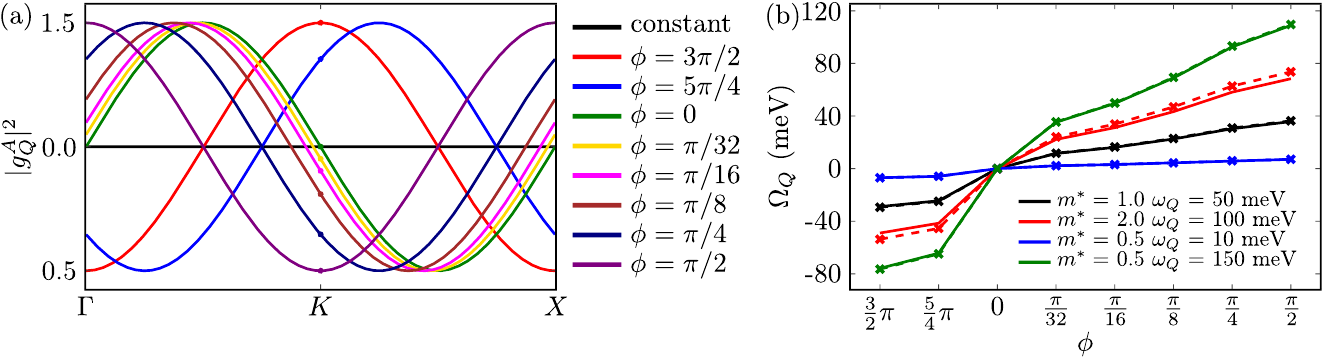}
    \caption{(a) $|g_{\rm k\twokf}(\phi)|^2$ as defined in Eq.~\eqref{eqSI:g2sin}, for different values of $\phi$. 
    (b) Renormalized phonon frequency $\Omega_{\rm \twokf}$ computed for different values of $m^*$ and $\omega_\twokf$ as a function $\phi$. 
    The solid lines are computed with $|g_{\rm k\twokf}(\phi)|^2$ as in Eq.~\eqref{eqSI:g2sin} while the dashed lines with crosses using $|g_{\rm k\twokf}(\phi)|^2 = |g_{\rm \kf\twokf}(\phi)|^2$.}
    \label{figSI:sinkG}
\end{figure}

First, we compute the renormalized frequency for different $|g_{\rm k\twokf}|^2$ values of the form:
\begin{equation}
    |g_{\rm k\rm \twokf}^\mathrm{A}(\phi)|^2 = 1+\frac{1}{2}\sin(2k+\phi),
    \label{eqSI:g2sin}
\end{equation}
where $|g_{\rm k\twokf}^\mathrm{A}(\phi)|^2$ is:
\begin{equation}
    |g_\twokf^\mathrm{A}(\phi)|^2 = \frac{|g_{\rm \twokf}(\phi)|^2}{<|g_{\rm \twokf}(\phi)|^2>},
    \label{eqSI:gA}
\end{equation}
where $<|g_{\rm \twokf}(\phi)|^2>$ indicates its average value.
In Fig.~\ref{figSI:sinkG}(a), Eq.~\eqref{eqSI:g2sin} is shown for various values of $\phi$ as a function of $k$, the value at $k=\kf$ is highlighted by a point.
In panel (b), the renormalized frequency is plotted for different values of $m^*$ and $\omega_\twokf$ using the following values of the EPC matrix elements: (i) the solid lines are computed using
\begin{equation}
    |g_{\rm k\twokf}(\phi)|^2 = g_0^2 |g_{\rm \twokf}^\mathrm{A}(\phi)|^2,
\end{equation}
with $g_0$ as defined in Eq.~\eqref{eq:g0} of the main manuscript, and (ii) the dashed lines with crosses are computed using 
\begin{equation}
    |g_{\rm k\twokf}(\phi)|^2 = |g_{\kf\twokf}(\phi)|^2.
\end{equation}
We notice that the two results are in agreement.

\begin{figure}[ht]
    \centering
    \includegraphics[width=0.99\linewidth]{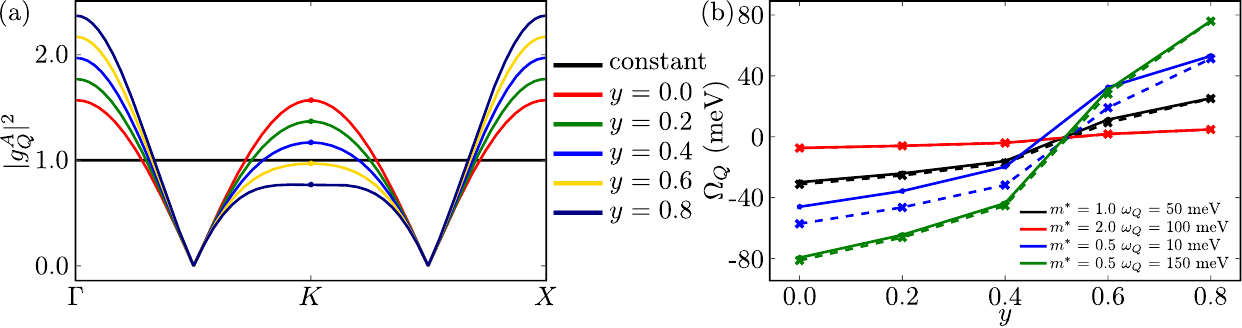}
    \caption{(a) $|g_{\rm k\twokf}(y)|^2$ as defined in Eq.~\eqref{eqSI:gp}, for different values of $y$ as a function of $k$. 
    (b) Renormalized phonon frequency $\Omega_{\rm \twokf}$ computed for different values of $m^*$ and $\omega_\twokf$ as a function of $y$. 
    The solid lines are computed using $|g_{\rm k\twokf}(y)|^2$ as in Eq.~\eqref{eqSI:gp} while the dashed lines with crosses using $|g_{\rm k\twokf}(y)|^2 = |g_{\rm \kf\twokf}(y)|^2$. }
    \label{figSI:gp}
\end{figure}

Another example is provided in Fig.~\ref{figSI:gp} where $|g_{\rm \twokf}^\mathrm{A}|^2$ is:
\begin{equation}
    |g_{\rm k\twokf}^\mathrm{A}(y)|^2 = |\cos(2k)|(\cos(2k)y+1.5693359375).
    \label{eqSI:gp}
\end{equation}
The offset $1.5693359375$ is chosen so that all curves have an average value of $1$.
Similarly to the previous case, in Fig.~\ref{figSI:gp}(b), the renormalized phonon frequency $\Omega_{\rm \twokf}$ is computed using
\begin{equation}
    |g_{\rm k\twokf}(\phi)|^2 = g_0^2 |g_{\rm \twokf}^\mathrm{A}(\phi)|^2,
\end{equation}
represented by solid lines, while the dashed lines with crosses correspond to computations using:
\begin{equation}
    |g_{\rm k\twokf}(\phi)|^2 = |g_{\kf\twokf}(\phi)|^2.
\end{equation}
We can notice once again a good agreement between the two computations.
Therefore, we conclude that when comparing the result of the model from this work with a real system, the value for comparison is that at the Fermi level.

\begin{figure*}[ht]
    \centering
    \includegraphics[width=0.85\linewidth]{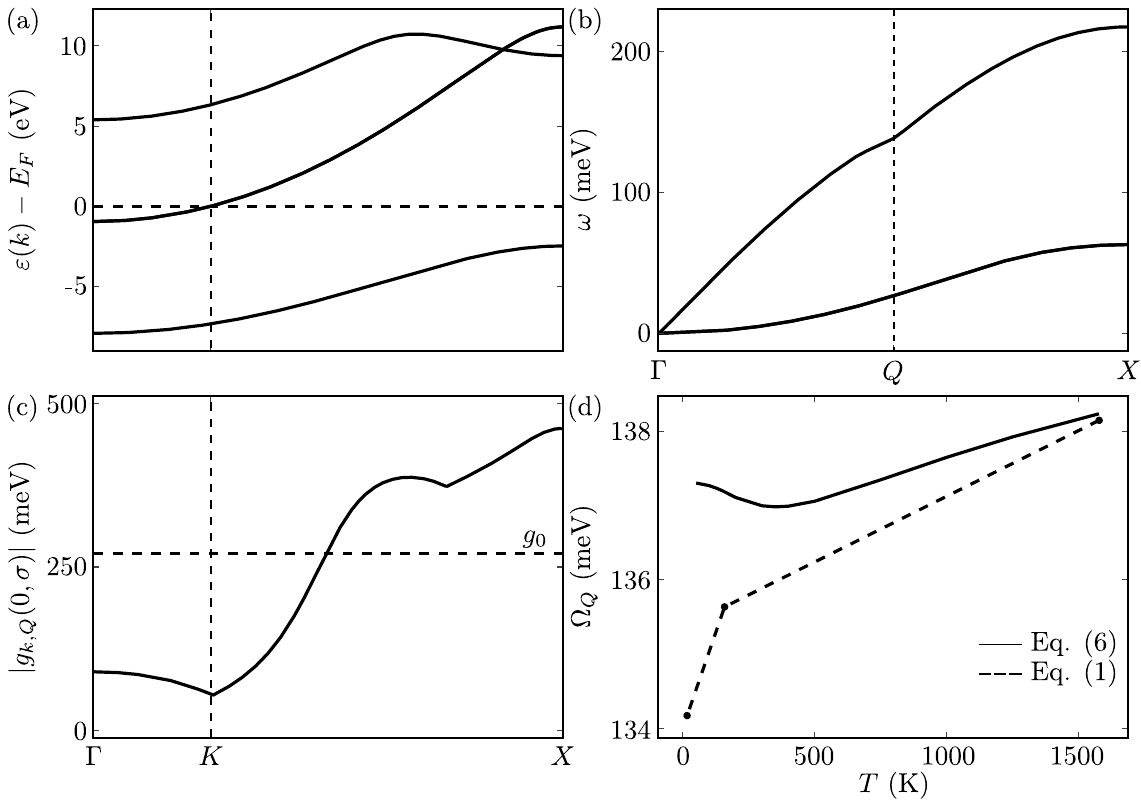}
    \caption{\label{fig:boron}Results for the boron atomic chain. 
    (a) Electron bandstructure and (b) phonon bandstructure, computed with a smearing parameter $\sigma=10$~mRy. 
    (c) $|g_{\rm k\twokf}^{nm\rm LA}|$ of the longitudinal acoustic mode. 
    The black line is the EPC matrix element computed using DFPT, while the dashed line is $g_0$ as defined in Eq.\eqref{eq:g0}. 
    (d) The renormalized phonon frequency of the longitudinal acoustic mode as a function of the temperature. 
    The solid curve is computed through Eq.~\eqref{eq:O_renorm_tot}, while the dashed curve corresponds to DFPT computations using a smearing of $\sigma$ = 10, 1 and, 0.1~mRy and plotted with $T=\sigma$.}
\end{figure*}

\section{Boron and strained gold monoatomic chains}\label{secSI:DFT}
In this section, the results for the boron and strained gold monoatomic chains are introduced.
They further validate the results of the theoretical model.

The monoatomic, evenly spaced boron chain is a metallic system with two degenerate $sp$-type bands that cross the Fermi level at $\kf=0.25\Gamma$-$X$ (see Fig.~\ref{fig:boron}(a)). 
This band structure suggests a perfect FS nesting at $\twokf=0.5$~$\Gamma$-$X$. 
The phonon bandstructure, computed with $\sigma=10$~mRy, is shown in Fig.~\ref{fig:boron}(b). 
A small renormalization appears for the longitudinal mode at $q = \twokf$.
Performing DFPT computation with different smearing parameters, a small dip is observed at $q=\twokf$, yet the frequency does not become imaginary regardless of how low the smearing parameter is set.
The dashed line in Fig.~\ref{fig:boron}(d) shows the DFPT frequencies with smearing parameters as low as 0.1~mRy.
To apply the results of the theoretical model to the longitudinal mode, we estimate the effective mass and the adiabatic phonon frequency. 
From the first-principles results, we determine $m^*\approx 0.67$ and $\omega_{\twokf} \approx 138.15$~meV, resulting in $g_0 \approx 281.44$~meV. 
Comparing this value with $|g_{\kf\twokf}^{nm\mathrm{LA}}|$ at the Fermi level, approximately 50~meV, we conclude that the expected renormalization is negligible and the system remains stable.
Although $|g_{\rm k\twokf}^{nm\mathrm{LA}}|$ is overall comparable to $g_0$, a good way to compare the two is to use the value at the Fermi level, as explained in Sec.~\ref{secSI:gkF2KF}.
Since $|g_{\kf\twokf}^{nm\rm LA}| \ll g_0$, we expect $\Omega_{\twokf}$ to remain close to $\omega_{\twokf}$ and to reach a minimum at approximately $T \approx 350$~K, using Eq.~\eqref{eq:Tmin} of the main manuscript.
In Fig.~\ref{fig:boron}(d), the renormalized phonon frequencies $\Omega_{\twokf}$ are shown.
The frequency decreases, reaching a minimum around 350~K as expected, then increases again and stabilizes at low temperature. 
Additionally, we observe that the renormalization is modest, as predicted, and that the renormalized frequency remains larger than the DFPT one.
This phenomenon has also been observed in similar computations, such as in Ref.~\cite{caruso2017}, where DFPT calculations were found to overestimate the effect of temperature on the phonon dispersion of boron-doped diamond. 
\begin{figure*}[ht]
    \centering
    \includegraphics[width=0.85\linewidth]{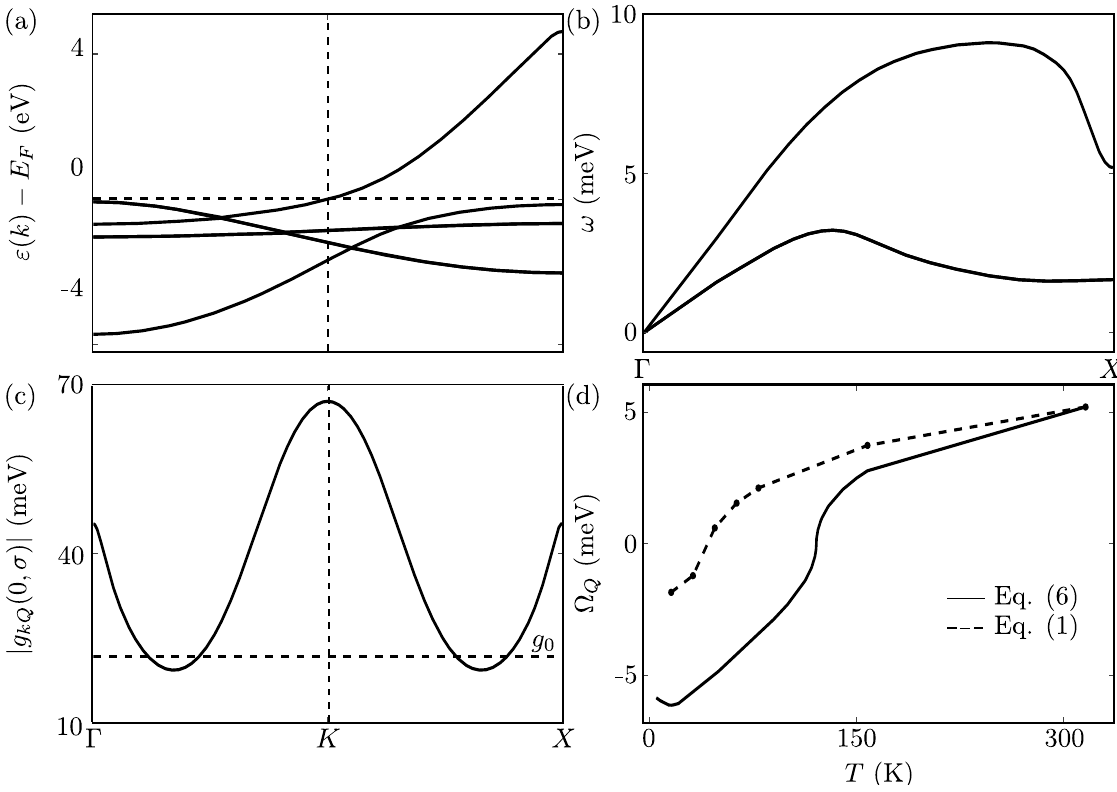}
    \caption{\label{fig:gold}Results for the strained gold atomic chain. 
    (a) Electron bandstructure and (b) phonon bandstructure computed with a smearing parameter $\sigma=2$~mRy. 
    (c) $|g_{\rm k\twokf}^{nm\rm LA}|$ of the longitudinal acoustic mode. 
    The black line is the EPC matrix element computed with DFPT, while the dashed line is $g_0$ as defined in Eq.~\eqref{eq:g0}. 
    (d) Renormalized phonon frequency of the longitudinal acoustic mode as a function of the temperature. 
    The black curve is computed though Eq.~\eqref{eq:O_renorm_tot}, while the dashed curve represents DFPT computations using a smearing of $\sigma$ = $2$, $1$, $0.5$, $0.4$, $0.3$, $0.2$ and, $0.1$~mRy and shown with $T=\sigma$.}
\end{figure*}

Finally, we analyze the strained gold chain~\cite{He2020}. 
The unstrained linear chain is unstable, but applying a strain of $\epsilon= 0.09$, where $\epsilon = d/d_0-1$ and $d_0 = 2.54$ \r{A}~\cite{He2020}, (i) stabilizes the transverse mode and (ii) fixes the position of the Fermi level at half of the $\Gamma$-$X$ line.
The DFT electron bandstructure presents a single $s$-type band crossing the Fermi level, the DFPT phonon bandstructure, computed at $\sigma=2$~mRy shows a dip at $q=\twokf=X$.
The two bandstructures are shown in Fig.~\ref{fig:gold}(a) and (b), respectively.  
At $q=\twokf$, the DFPT longitudinal-acoustic phonon frequency is approximately $5.19$~meV and it rapidly becomes imaginary as the smearing parameter decreases further.
In Fig.~\ref{fig:gold}(d), the dashed line represents the DFPT computations performed with smearing values $\sigma$ ranging from $2$ to $0.1$~mRy reported as a fictitious temperature $T$.
We note that the DFPT transition temperature is around $30$~K.
This result is consistent with the findings of Ref.~\cite{He2020} where, using the Fermi-Dirac smearing scheme, they obtained imaginary frequencies around $0.1$~mHa, corresponding to $30$~K. 
Once again, to apply the results of the theoretical model to the longitudinal mode, we estimate the effective mass and the adiabatic phonon frequency. 
The adiabatic phonon frequency is $\omega_{\rm \twokf}\approx 5.19$~meV, while $m^*\approx3$.
With these values, we compute $g_0\approx17.6$~meV.
The EPC matrix elements computed using DFPT and $g_0$ are shown in Fig.~\ref{fig:gold}(c) with solid and dashed lines, respectively.
We notice that $|g_{\rm k\twokf}^{nm\nu}|$ exceeds $g_0$.
Therefore, we expect the renormalized frequency to become imaginary.
In Fig.~\ref{fig:gold}(d), the renormalized frequency $\Omega_{\twokf}$ is plotted.
The plot shows a sharp transition to imaginary frequencies around 120~K, refining the value found in Ref.~\cite{He2020}.
In the case of the gold chain, the smaller bare phonon frequency, larger effective mass, and stronger electron-phonon coupling lead to a more pronounced renormalization, ultimately driving the frequency to become imaginary.
As for the case of the RBM in CNT (3,3), shown in the main manuscript, when the renormalization leads to an imaginary frequency, the DFPT computation underestimates the effect of the EPC.
In contrast, when the renormalized frequency remains real, as in the case of boron and $A_1(T)$ and $A_1(L)$ modes in CNT (3,3) and the boron chain, the DFPT calculation overestimates the effect of EPC.

\bibliographystyle{unsrt}
\bibliography{bib}

\begin{thebibliography}{10}

\bibitem{kohn1959}
W.~Kohn.
\newblock Image of the fermi surface in the vibration spectrum of a metal.
\newblock {\em Phys. Rev. Lett.}, 2:393--394, 1959.

\bibitem{brockhouse1961}
B.~N. Brockhouse, K.~R. Rao, and A.~D.~B. Woods.
\newblock Image of the fermi surface in the lattice vibrations of lead.
\newblock {\em Phys. Rev. Lett.}, 7:93--95, 1961.

\bibitem{brockhouse1962}
B.~N. Brockhouse, T.~Arase, G.~Caglioti, K.~R. Rao, and A.~D.~B. Woods.
\newblock Crystal dynamics of lead. i. dispersion curves at
  100\ifmmode^\circ\else\textdegree\fi{}k.
\newblock {\em Phys. Rev.}, 128:1099--1111, 1962.

\bibitem{nakagawa1963}
Y.~Nakagawa and A.~D.~B. Woods.
\newblock Lattice dynamics of niobium.
\newblock {\em Phys. Rev. Lett.}, 11:271--274, 1963.

\bibitem{koenig1964}
S.~H. Koenig.
\newblock Kohn effect in na and other metals.
\newblock {\em Phys. Rev.}, 135:A1693--A1695, 1964.

\bibitem{baron2004}
A.~Q.~R. Baron, H.~Uchiyama, Y.~Tanaka, S.~Tsutsui, D.~Ishikawa, S.~Lee,
  R.~Heid, K.-P. Bohnen, S.~Tajima, and T.~Ishikawa.
\newblock Kohn anomaly in ${\mathrm{mgb}}_{2}$ by inelastic x-ray scattering.
\newblock {\em Phys. Rev. Lett.}, 92:197004, 2004.

\bibitem{aynajin2008}
P.~Aynajian, T.~Keller, L.~Boeri, S.~M. Shapiro, K.~Habicht, and B.~Keimer.
\newblock Energy gaps and kohn anomalies in elemental superconductors.
\newblock {\em Science}, 319(5869):1509--1512, 2008.

\bibitem{powell1968}
B.~M. Powell, P.~Martel, and A.~D.~B. Woods.
\newblock Lattice dynamics of niobium-molybdenum alloys.
\newblock {\em Phys. Rev.}, 171:727--736, 1968.

\bibitem{kulda2002}
J.~Kulda, H.~Kainzmaier, D.~Strauch, B.~Dorner, M.~Lorenzen, and M.~Krisch.
\newblock Overbending of the longitudinal optical phonon branch in diamond as
  evidenced by inelastic neutron and x-ray scattering.
\newblock {\em Phys. Rev. B}, 66:241202, 2002.

\bibitem{peierls1955}
R.~E. Peierls.
\newblock {\em Quantum Theory of Solids}.
\newblock Oxford University Press, 1955.

\bibitem{zhu2015}
X.~Zhu, Y.~Cao, J.~Zhang, E.~W. Plummer, and J.~Guo.
\newblock Classification of charge density waves based on their nature.
\newblock {\em PNAS}, 112:2367--2371, 2015.

\bibitem{johannes2008}
M.~D. Johannes and I.~I. Mazin.
\newblock Fermi surface nesting and the origin of charge density waves in
  metals.
\newblock {\em Phys. Rev. B}, 77:165135, 2008.

\bibitem{tidholm2020}
J.~Tidholm, O.~Hellman, N.~Shulumba, S.~I. Simak, F.~Tasn\'adi, and I.~A.
  Abrikosov.
\newblock Temperature dependence of the kohn anomaly in bcc nb from
  first-principles self-consistent phonon calculations.
\newblock {\em Phys. Rev. B}, 101:115119, 2020.

\bibitem{berges2023}
J.~Berges, N.~Girotto, T.~Wehling, N.~Marzari, and S.~Ponc\'e.
\newblock Phonon self-energy corrections: To screen, or not to screen.
\newblock {\em Phys. Rev. X}, 13:041009, 2023.

\bibitem{caruso2017}
F.~Caruso, M.~Hoesch, P.~Achatz, J.~Serrano, M.~Krisch, E.~Bustarret, and
  F.~Giustino.
\newblock Nonadiabatic kohn anomaly in heavily boron-doped diamond.
\newblock {\em Phys. Rev. Lett.}, 119:017001, 2017.

\bibitem{girotto2023}
N.~Girotto and D.~Novko.
\newblock Dynamical renormalization of electron-phonon coupling in conventional
  superconductors.
\newblock {\em Phys. Rev. B}, 107:064310, 2023.

\bibitem{gonze1997}
X.~Gonze and C.~Lee.
\newblock Dynamical matrices, born effective charges, dielectric permittivity
  tensors, and interatomic force constants from density-functional perturbation
  theory.
\newblock {\em Phys. Rev. B}, 55:10355--10368, 1997.

\bibitem{gonze1997a}
X.~Gonze.
\newblock First-principles responses of solids to atomic displacements and
  homogeneous electric fields: Implementation of a conjugate-gradient
  algorithm.
\newblock {\em Phys. Rev. B}, 55:10337--10354, 1997.

\bibitem{Calandra2010}
M.~Calandra, G.~Profeta, and F.~Mauri.
\newblock Adiabatic and nonadiabatic phonon dispersion in a wannier function
  approach.
\newblock {\em Physical Review B}, 82(16), October 2010.

\bibitem{giustino2017}
F.~Giustino.
\newblock Electron-phonon interactions from first principles.
\newblock {\em Rev. Mod. Phys.}, 89:015003, 2017.

\bibitem{Marini2025}
A.~Marini.
\newblock Dynamical electron-phonon vertex correction, 2025.

\bibitem{Stefanucci2025}
G.~Stefanucci and E.~Perfetto.
\newblock Exact formula with two dynamically screened electron-phonon couplings
  for positive phonon-linewidths approximations.
\newblock {\em Physical Review B}, 111(2), January 2025.

\bibitem{SI}
See Supplemental Material at [URL] for details on the 1D model, the analitical
  computation of polarizability at T = 0 K, its temperature dependent
  approximations, details on the EPC matrix element value to compare with
  theory and other first-principle examples, which includes
  Refs.~\cite{caruso2017,He2020}.

\bibitem{QE-2009}
P.~Giannozzi, S.~Baroni, N.~Bonini, M.~Calandra, R.~Car, C.~Cavazzoni,
  D.~Ceresoli, G.~L. Chiarotti, M.~Cococcioni, I.~Dabo, and et~al.
\newblock Quantum espresso: a modular and open-source software project for
  quantum simulations of materials.
\newblock {\em J Phys: Condens Matter}, 21(39):395502, 2009.

\bibitem{QE-2017}
P~Giannozzi, O~Andreussi, T~Brumme, O~Bunau, M~Buongiorno Nardelli, M~Calandra,
  R~Car, C~Cavazzoni, D~Ceresoli, and et~al.
\newblock Advanced capabilities for materials modelling with quantum espresso.
\newblock {\em J Phys: Condens Matter}, 29(46):465901, 2017.

\bibitem{QE-2020}
P.~Giannozzi, O.~Baseggio, P.~Bonfà, D.~Brunato, R.~Car, I.~Carnimeo,
  C.~Cavazzoni, S.~de~Gironcoli, P.~Delugas, F.~Ferrari~Ruffino, A.~Ferretti,
  N.~Marzari, I.~Timrov, A.~Urru, and S.~Baroni.
\newblock Quantum espresso toward the exascale.
\newblock {\em J. Chem. Phys.}, 152(15):154105, 2020.

\bibitem{Giustino2007}
F.~Giustino, M.~L. Cohen, and S.~G. Louie.
\newblock Electron-phonon interaction using wannier functions.
\newblock {\em Phys. Rev. B}, 76:165108, 2007.

\bibitem{Ponce2016}
S.~Ponc\'e, E.R. Margine, C.~Verdi, and F.~Giustino.
\newblock Epw: Electron-phonon coupling, transport and superconducting
  properties using maximally localized wannier functions.
\newblock {\em Computer Physics Communications}, 209:116 -- 133, 2016.

\bibitem{Lee2023}
H.~Lee, S.~Ponc\'e, K.~Bushick, S.~Hajinazar, J.~Lafuente-Bartolome,
  J.~Leveillee, C.~Lian, J.~Lihm, F.~Macheda, H.~Mori, and et~al.
\newblock Electron--phonon physics from first principles using the epw code.
\newblock {\em npj Computational Materials}, 9:156, 2023.

\bibitem{vansetten2018}
M.J. van Setten, M.~Giantomassi, E.~Bousquet, M.J. Verstraete, D.R. Hamann,
  X.~Gonze, and G.-M. Rignanese.
\newblock The pseudodojo: Training and grading a 85 element optimized
  norm-conserving pseudopotential table.
\newblock {\em Comput Phys Commun}, 226:39--54, 2018.

\bibitem{Marzari2012}
N.~Marzari, A.~A. Mostofi, J.~R. Yates, I.~Souza, and D.~Vanderbilt.
\newblock Maximally localized wannier functions: Theory and applications.
\newblock {\em Rev. Mod. Phys.}, 84:1419--1475, 2012.

\bibitem{w90}
G.~Pizzi, V.~Vitale, R.~Arita, S.~Blügel, F.~Freimuth, G.~Géranton,
  M.~Gibertini, D.~Gresch, C.~Johnson, and T.~Koretsune.
\newblock Wannier90 as a community code: new features and applications.
\newblock {\em J Phys: Condens Matter}, 32:165902, 2020.

\bibitem{connetable2005}
D.~Connétable, G.-M. Rignanese, J.-C. Charlier, and X.~Blase.
\newblock Room temperature peierls distortion in small diameter nanotubes.
\newblock {\em Phys. Rev. Lett.}, 94:015503, 2005.

\bibitem{barnett2005}
R.~Barnett, E.~Demler, and E.~Kaxiras.
\newblock Electron-phonon interaction in ultrasmall-radius carbon nanotubes.
\newblock {\em Phys. Rev. B}, 71:035429, 2005.

\bibitem{marzari1999}
N.~Marzari, D.~Vanderbilt, A.~De~Vita, and M.~C. Payne.
\newblock Thermal contraction and disordering of the al(110) surface.
\newblock {\em Phys. Rev. Lett.}, 82:3296--3299, 1999.

\bibitem{bohnen2004}
K.-P. Bohnen, R.~Heid, H.~J. Liu, and C.~T. Chan.
\newblock Lattice dynamics and electron-phonon interaction in (3,3) carbon
  nanotubes.
\newblock {\em Phys. Rev. Lett.}, 93:245501, 2004.

\bibitem{He2020}
L.~He, F.~Liu, J.~Li, G.-M. Rignanese, and A.~Zhou.
\newblock First-principles investigation of monatomic gold wires under tension.
\newblock {\em Comput Mater Sci}, 171:109226, 2020.

\end{thebibliography}

\end{document}